\documentclass[%
 reprint,
 amsmath,amssymb,
 aps,
 prc,
]{revtex4-1}

\usepackage{graphicx}
\usepackage{dcolumn}
\usepackage{bm}
\usepackage{lineno}
\usepackage{multirow}

\begin{document}

\preprint{APS/123-QED}

\title{Measurement of absorption and charge exchange of $\pi^+$ on carbon}

\author{K. Ieki$^1$, E. S. Pinzon Guerra$^2$, S. Berkman$^3$, S. Bhadra$^2$, C. Cao$^3$, P. de Perio$^4$, Y. Hayato$^5$, \\M. Ikeda$^5$, Y. Kanazawa$^6$,  J. Kim$^3$, P. Kitching$^7$, K. Mahn$^8$, T. Nakaya$^1$,  M. Nicholson$^7$, K. Olchanski$^7$, S. Rettie$^{3,7}$, H. A. Tanaka$^3$, S. Tobayama$^3$, M. J. Wilking$^9$, T. Yamauchi$^1$, S. Yen$^7$, M. Yokoyama$^6$}

\affiliation{%
$^1$Kyoto University, Department of Physics, Kyoto, Japan\\
$^2$York University, Department of Physics and Astronomy, Toronto, Ontario, Canada\\
$^3$University of British Columbia, Department of Physics and Astronomy, Vancouver, British Columbia, Canada\\
$^4$University of Toronto, Department of Physics, Toronto, Ontario, Canada\\
$^5$University of Tokyo, Institute for Cosmic Ray Research, Kamioka Observatory, Kamioka, Japan\\
$^6$University of Tokyo, Department of Physics, Tokyo, Japan\\
$^7$TRIUMF, Vancouver, British Columbia, Canada\\
$^8$Michigan State University, Department of Physics and Astronomy, East Lansing, Michigan, U.S.A\\
$^9$State University of New York at Stony Brook, Department of Physics and Astronomy, Stony Brook, New York, U.S.A.
}%

\collaboration{DUET Collaboration}



\date{\today}

\begin{abstract}

  The combined cross section for absorption and charge exchange interactions of positively charged pions 
  with carbon nuclei for the momentum range 200 MeV/c to 300 MeV/c have been measured with the DUET experiment
  at TRIUMF. 
  The uncertainty is reduced by nearly half compared to 
  previous experiments. This result will be a valuable input to existing models
  to constrain pion interactions with nuclei.

\end{abstract}
\pacs{Valid PACS appear here}
\maketitle


\section{\label{sec:intro}Introduction\protect\\}


It is widely believed that strong interactions are governed by quantum
chromodynamics (QCD), which implies that the structure of both atomic
nuclei and their constituent nucleons are fully described by the
interactions of quarks and gluons. However, at separation distances
typically found between the nucleons within an atomic nucleus
($\sim$1~fm), color confinement suggests that the interactions between
nucleons can be described by the exchange of colorless
particles. Effective theories based on the interactions between nucleons
and mesons can therefore be constructed to describe nuclear structure,
and such interactions can be directly probed by experiments that study
the scattering of pions off of atomic nuclei.

\begin{figure}[htbp]
 \begin{center}
  \includegraphics[width=90mm]{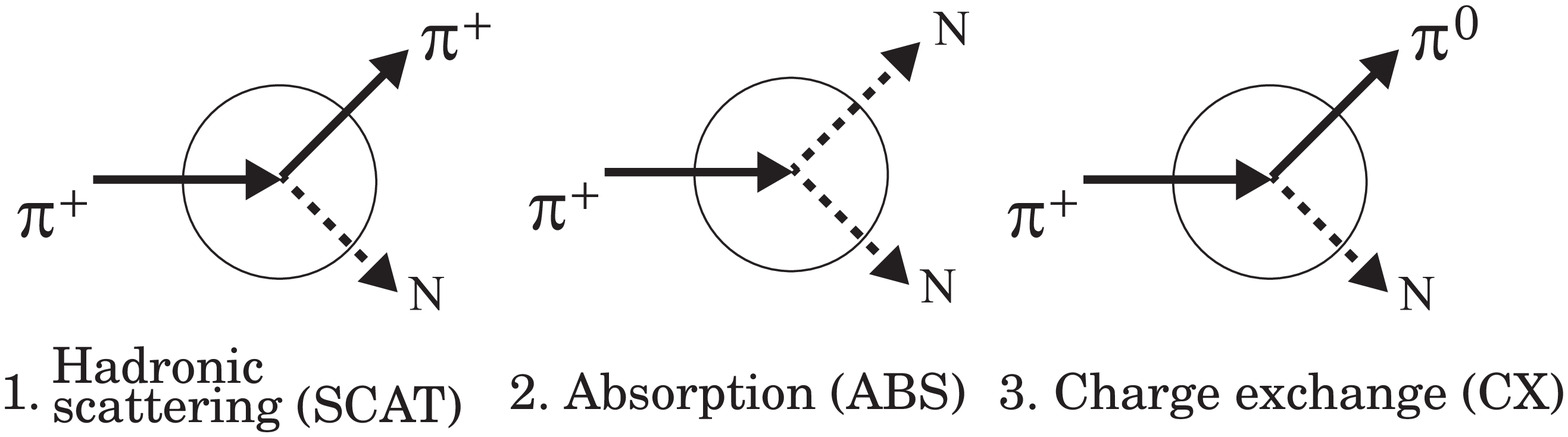}
 \end{center}
 \caption{Pion interactions on nuclei. ``N'' represents any number of nucleons emitted after interactions }
 \label{fig:pi_intr}
\end{figure}

Over the past forty years, an extensive set of pion scattering
experiments have been conducted at various meson factories, such as the
Los Alamos Meson Physics Facility (LAMPF) in the United States, the Paul
Scherrer Institute (PSI) in Switzerland, and the TRIUMF laboratory in
Canada~\cite{Rowntree,Giannelli,Nakai,Saunders,Navon,Bowles,Carroll,Ashery,Ashery2,Clough,Levenson,Ingram,Wood,Miller,Jones,Gelderloos,Crozon,Takahashi,Allardyce,Cronin,Fujii,Aoki,Grion,Rahav}.
Although these data have provided very detailed measurements of
differential cross sections for a variety of final state kinematic
variables, the uncertainties on the inclusive cross sections for
processes such as pion absorption and charge exchange (see Figure~\ref{fig:pi_intr}) range from 10-30\%
for light nuclei, such as carbon and oxygen. Of particular interest are
pion absorption measurements, in which an incident $\pi^+$ interaction
fails to produce a pion in the final state. Since a pion cannot be
absorbed by a nucleon in a manner that conserves energy and momentum,
absorption interactions must involve coupled states of at least two
nucleons. Pion absorption measurements therefore provide unique insight
into nuclear structure by directly probing the correlations between
component nucleons.

Beyond intrinsic theoretical interest in nuclear structure, pion
interactions can play a critical role in understanding systematic
uncertainties in experiments conducted at the GeV energy scale. One such
field that is sensitive to pion cross section uncertainties is the study
of neutrinos. When a neutrino interacts with an atomic nucleus via
a charged current interaction, a charge lepton is produced.
In experiments studying the interactions of neutrinos with incident 
energy around 1 GeV, the energy of the neutrino is typically inferred from
the measured kinematics of the outgoing lepton and the assumed recoil mass
of the target nucleon. Around this energy, the cross section for neutrino-induced
pion production is large.
If pions is produced, but not detected
due to 
interactions within the target nucleus or 
after exiting the nucleus, the inferred neutrino energy will be
biased. Pions with momenta of a few hundred MeV/c interact primarily 
in three modes as shown in Figure~\ref{fig:pi_intr}: 1) Hadronic scattering
through inelastic (quasi-elastic) and elastic channels (SCAT), 
2) Absorption (ABS) and 3) Charge exchange (CX).
Interactions in which a $\pi^{\pm}$ does not produce a
$\pi^{\pm}$ such as pion absorption and charge exchange interactions
can be particularly challenging to reconstruct, since low energy
nucleons and photons from $\pi^0$ decay can be difficult to detect. 
The contribution of double charge exchange interaction is small for light nuclei.

The Dual-Use Experiment at TRIUMF (DUET) is intended to improve the
precision of pion absorption and charge exchange interaction cross 
sections on both carbon and water.
A scintillator tracker is used for precision studies of 
pion interaction final states.
The experiment is capable of measuring interactions on carbon and water.
A limited number of photon detectors were deployed to allow the
separation of absorption and charge exchange interactions. In this paper,
we present a measurement of the combined absorption and charge exchange
cross section ($\sigma_{\mathrm{ABS}+\mathrm{CX}}$) on carbon with significantly improved precision relative to previous measurements.

 \section{\label{sec:experiment}Experiment\\}

 $\sigma_{\mathrm{ABS}+\mathrm{CX}}$ was measured 
 on a carbon target at 5 different momentum settings between 201.6 MeV/c and 295.5 MeV/c.
 The ABS and CX events are selected by requiring no observed pion in the final state, 
 therefore a detector with excellent tracking capabilities was essential.
 The pion interactions were measured within the PIA$\nu$O (PIon
 detector for Analysis of neutrino Oscillation) detector, which 
 was composed of 1.5 mm scintillating fibers to provide precise tracking and 
 dE/dx measurements of particles in the interaction final state.



  
  \begin{figure}[!h]
   \begin{center}
    \includegraphics[width=90mm]{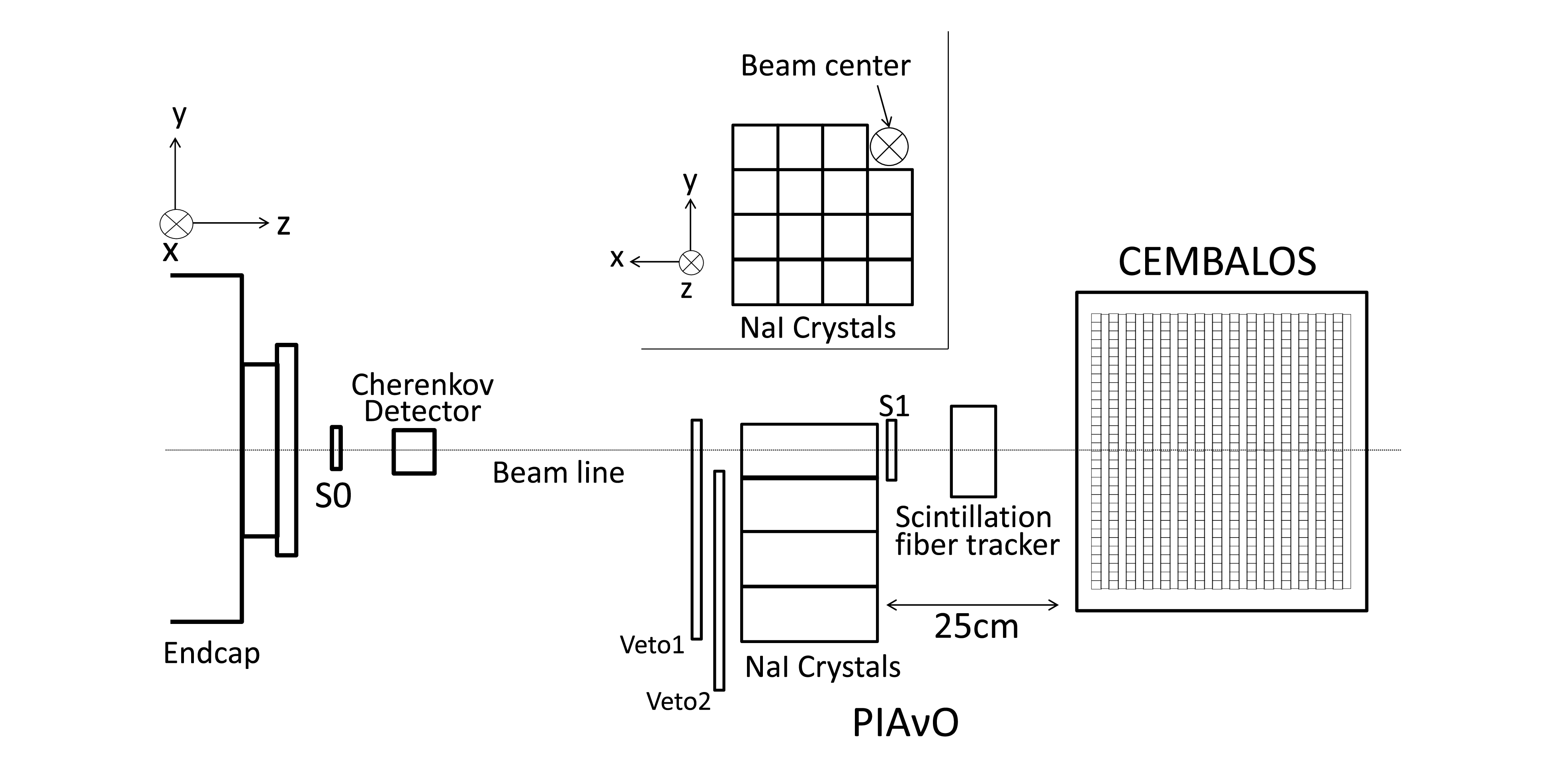}
    \caption{Apparatus layout. Detailed description in the text.\label{fig:config}}
   \end{center} 
  \end{figure}




  \subsection{Beam line and triggers}\label{sec:beamline}
  The experiment took place in the M11 secondary beam line at TRIUMF. 
  Figure~\ref{fig:config} shows the overview of the M11 beamline area and
  the placement of the detectors. 
  A 500 MeV proton beam extracted from the TRIUMF main cyclotron was directed onto a 1 cm carbon target. 
  The pions produced in the target were directed down the M11 beam line 
  by two dipole magnets and focused by a series of six quadrupole magnets.  
  The accelerator facility allowed the possibility to select different pion momenta, 
  and the momentum settings used were 201.6, 216.6, 237.2, 265.5, 295.1 MeV/c.
  The momentum of the pion beam is measured using CEMBALOS,
  described in Chapter \ref{sec:xsec_calc}.
  
\begin{figure}[htbp]
	\begin{center}
	 \includegraphics[width=85mm]{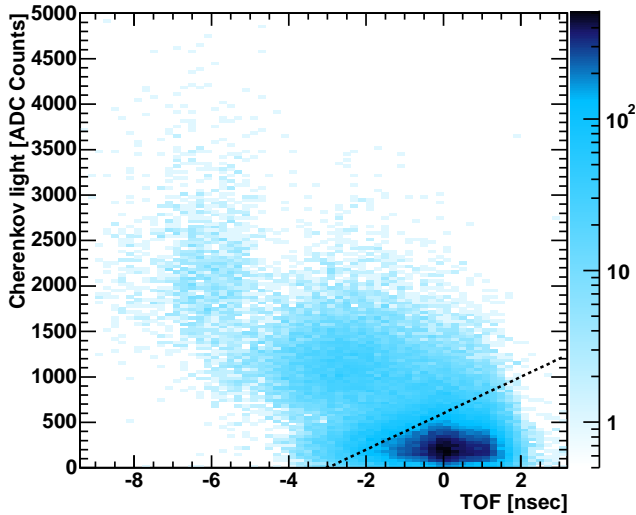}
	\end{center}
	\caption{
	Cherenkov light in ADC counts vs. TOF [nsec] for the beam particle at $p_\pi=237.2$ MeV/c
	setting. 
	The broken line corresponds to the threshold to distinguish pions
	from muons and electrons.
	}
	\label{fig:tof_cher}
       \end{figure}

    In addition to pions, the secondary beam also contained protons produced
    from the target, and muons and electrons resulting from the pion decay
    chain. 
    The pions were selected using Time-Of-Flight (TOF) measurements and a Cherenkov detector. 
    The TOF of each secondary particle was the difference between the time measured in the Current 
    Transformer (CT), located near the production target, and scintillator counter
    S1, placed $\sim$15 m downstream of the CT.
    The CT, S0, and S1 detectors were read out by a VME module (CAEN TDC V1190), 
    and the TOF determined by the difference in TDC counts between S1 and the CT.
    A Cherenkov counter was placed $\sim$11 cm downstream of the S0 counter,
    and consisted of a 3.5 cm $\times$ 3.5 cm $\times$ 20 cm bar of
    Bicron UV-transparent acrylic plastic read out at each end by 
    photomultiplier tubes. The refractive index
    of the acrylic bar was 1.49, so muons with momentum larger than
    $\sim$250 MeV/c produced Cherenkov light at angles that 
    were totally internally reflected within the bar, whereas pions of the 
    same momentum produced Cherenkov light at an angle that was
    largely transmitted. 
    The signals of the two PMTs were read out by a VME module (CAEN ADC
    V792), and the Cherenkov light for each event was obtained 
    from the sum of the ADC
    counts of the two PMTs. 
    Figure \ref{fig:tof_cher} shows an example of
    Cherenkov light vs. TOF for $p_\pi=$ 237.2 MeV/c. The
    electron, muon and pion signals are clustered around the left 
    top, middle and right bottom of the plot, respectively. The pion
    candidates are below the broken line. 
    The purity of pions after this cut is estimated to be larger than 99\% for all the
    momenta settings used in the analysis.

    The S0 and S1 scintillator counters were used in coincidence to select 
    low angle charged particles entering the PIA$\nu$O detector.
    

    

  \subsection{Detector description}

\begin{figure}[htbp]
   \begin{center}
    \includegraphics[width=85mm]{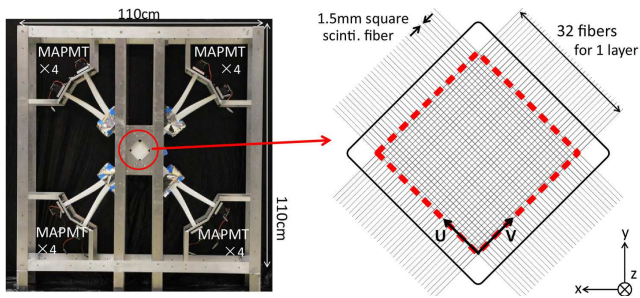}
   \end{center}
   \caption{
   Front view of the fiber tracker detector.
   }
   \label{fig:fiber_front}
  \end{figure}

  The PIA$\nu$O fiber tracker consists of 1.5 mm 
  scintillation fibers and is read out by Multi-Anode Photo Multiplier 
  Tubes (MAPMTs). Figure \ref{fig:fiber_front} shows the front view of
  the detector. The pion beam is injected into the center of the detector,
  where the fibers cross each other perpendicularly to form U and V
  layers. There are 16 U and 16 V layers, with 32 fibers in each layer
  for a total of 1024 fibers or channels. The dimension of the region
  where the fibers cross each other (``fiber crossing region'') is
  $\sim5\times5\times5$ cm$^3$. The fibers are held together
  by fiber holders to clip the fibers without glue. The fiber channels
  are read out by 16 MAPMTs.
  The structure of the detector, details of the fiber scintillators, the 
  MAPMTs, and the readout electronics are summarized in Table \ref{tb:spec_fiber}.

  \begin{table*}[h]
   \begin{center}
    \begin{tabular}{cc}
     \noalign{\hrule height 2pt}
     \multicolumn{2}{c}{Structure}\\
     \noalign{\hrule height 1pt}
     Dimensions in fiber crossing region & 49 mm $\times$ 49 mm $\times$ 51 mm\\
     Dimensions of support structure & 110 cm $\times$ 110 cm $\times$ 25 cm\\
     Number of channels & 1024\\
     \noalign{\hrule height 1pt}
     \multicolumn{2}{c}{Scintillating fiber}\\
     \noalign{\hrule height 1pt}
     Material & Polystyrene (core), PMMA (clad)\\
     Reflector & EJ-510 ($\sim$ 25 $\mu$m)\\
     Dimensions & 0.149 cm $\times$ 0.149 cm $\times$ 60 cm (core + clad)\\
     Clad thickness  & 2\% of core + clad \\
     Emission peak wavelength & 450 nm \\
     Decay time & 2.8 ns \\
     Attenuation length & $>$ 4 m \\
     \noalign{\hrule height 1pt}
     \multicolumn{2}{c}{MAPMT}\\
     \noalign{\hrule height 1pt}
     Type & Hamamatsu H8804\\
     Anode & 8$\times$8 pixels (pixel size: 2$\times$2 mm$^2$) \\
     Cathode & Bialkali (Sb-K-Cs) \\
     Sensitive wavelength & 300-650 nm (peak: 420 nm) \\
     Quantum efficiency & 12\% at $\lambda$=500 nm\\
     Dynode & Metal channel structure 12 stages\\
     Gain & typical 2 $\times$ 10$^6$ at 900 V \\
     Crosstalk & $\sim$2\% (adjacent pixel) \\
     \noalign{\hrule height 1pt}
     \multicolumn{2}{c}{Readout electronics}\\
     \noalign{\hrule height 1pt}
     Number of ADC channels & 1024\\
     ADC pedestal width & less than 0.1 p.e.\\
     \noalign{\hrule height 2pt}
    \end{tabular}
   \end{center}
   \caption{Specifications of the fiber tracker}	
   \label{tb:spec_fiber}
  \end{table*}

  The scintillating fibers used are single clad square
  fibers (Kuraray SCSF-78SJ). The outer surface of the fibers are coated with a 
  reflective coating (EJ-510) which contains TiO$_2$ to
  increase the light yield by trapping the light within the fiber and to optically 
  separate the fibers from each other.
  One end of each fiber is mirrored by vacuum deposition of
  aluminum which increases by 70\% the light yield. The number of nuclei in the fiducial volume of the fiber is estimated from
  the material and dimension of the fibers, as summarized in Table
  \ref{tbl:fib_nucl}.
  \begin{table*}[h]
   \begin{center}
    \begin{tabular}{cc}
     \noalign{\hrule height 2pt}
     Nuclei & Number of nuclei [$\times10^{24}$] \\ \hline
     C & 1.518$\pm$0.007 \\ \hline
     H & 1.594$\pm$0.008 \\\hline
     O & 0.066$\pm$0.004 \\\hline
     Ti & 0.006$\pm$0.0002 \\
     \noalign{\hrule height 2pt}
    \end{tabular}
    \caption{
    The number of nuclei in the fiducial volume of the fiber tracker.
    }
    \label{tbl:fib_nucl}
   \end{center}
  \end{table*}


  The scintillation light from the fibers is read out by 64 channel
  MAPMTs which are connected via acrylic connectors. A small fraction of
  the light from the fibers is transferred to adjacent MAPMT channels
  which generates crosstalk signals.
  Adjacent fibers in a layer are connected to non-adjacent MAPMT channels 
  so that crosstalk signals can be
  separated from the real signal. The crosstalk
  probability is measured to be $\sim$2\% for the adjacent 
  channels. The readout electronics for MAPMTs is recycled from the 
  K2K experiment\cite{scibar2}. 
  Each of the MAPMTs are read out through a front-end board. 
  Signals from the front-end board are digitized by Flash Analog to Digital Converters (FADCs) 
  on the back-end modules mounted on a VME-9U crate. 

  The high voltage for MAPMTs is tuned in a bench test by measuring 1
  photoelectron (p.e.) signals from LED light so that the gain 
  is uniform over all MAPMTs. 
  The high voltage is set to $\sim$950 V, and the typical gain is 60
  ADC/p.e. However, it varies by $\sim$23\% between MAPMT channels
  because the gain of the 64 channels within a MAPMT cannot be tuned
  individually. The measured light yield is $\sim$11 p.e. per fiber for
  a minimum ionizing particle.   
  The relatively large value of the MAPMT high voltage is necessary to measure the light from the
  fibers with good resolution. The dynamic range of FADCs is therefore
  not wide (maximum $\sim$30 p.e.).




	Using only the tracker, $\pi^0$s from charge exchange events cannot
	be observed. NaI detectors surrounding the tracker were installed to
	detect $\gamma$s from the decay of $\pi^0$s for separation of absorption
	and charge exchange events. The apparatus configuration also includes the 
	CEMBALOS (Charge Exchange Measurement By A Lead On
	Scintillator) detector. This was a scaled-down version of the T2K
	Fine-Grained Detectors (FGDs)~\cite{fgdnim},  with removable
	lead plates sandwiched between scintillator tracking planes to act as
	another photon detector, together with the NaI detectors. For this
	analysis, CEMBALOS was used for the evaluation of the systematics
	for the muon contamination of the beam. The NaI array and
	CEMBALOS are used in ongoing studies to extract ABS and CX cross
	sections separately and will be the subject of another paper. 

  \subsection{Detector Simulation}\label{sec:detector_sim}

  The detector simulation includes a detailed description of the tracker,
  Cherenkov counter, scintillator counters, and CEMBALOS.
  The simulation code is based on Geant4 version 9.4 patch 04~\cite{geant}.
  The fiber core, cladding and coating structure of PIA$\nu$O
  are included in the simulation. The thickness
  of the coating affects the efficiency to detect a hit above 2.5
  p.e. threshold for 
  through-going pions. The efficiency is measured to be $\sim$94\% in MC,
  while it is measured to be 93\% in the data. 

  The misalignment of the fiber layer position is
  measured from the difference between the measured hit position and the
  expected hit position for through-going pions. The RMS of the distance from
  the nominal position is measured to be 80 $\mu m$. The shift is implemented in the simulation by shifting the layer position to the
  measured position for that layer. The light yield of the fibers in the simulation is tuned so
  that it agrees with pion through-going data. 

  The energy deposit for
  each fiber in the simulation is converted to p.e. by the following
  procedure.
  \begin{enumerate}
   \item Conversion of energy deposit to photons\\
	 The expected number of photons generated in the fiber  ($N_{exp}$) is
	 calculated by multiplying the value of the energy deposit ($E_{dep}$) by a
	 conversion factor, $C_{conv}$ ($\sim$57 p.e./MeV), which is 
	 defined channel by channel from the light yield
	 distribution observed in through-going pion data. Thus\
	 \begin{equation}
	  N_{exp} = C_{conv} \times E_{dep} 
	 \end{equation} 
	 The saturation of scintillation light is
	 taken into account by using Birk's formula~\cite{Birk}.
         The Birk's constant for our fiber material (polystyrene) is the same as for the FGD~\cite{fgdnim}.

   \item Photon statistics  and MAPMT gain\\
         The photon statistics and the MAPMT gain fluctuation is taken into account. 
	 The number of photoelectrons ($N_{P.E.}$) is randomly defined from the
	 Poisson distribution using the mean of the expected number of photons 
         ($N_{P.E.} = \mbox{Poisson(}N_{exp}$)).
	 The observed number of photoelectrons ($N_{obs}$) is obtained
	 by adding a statistical fluctuation term to $N_{P.E.}$:
	 \begin{equation}
	  N_{obs} = N_{P.E.} + \sqrt{N_{P.E.}} \times C_{gain} \times \mbox{Gauss(1)} 
	 \end{equation} 
	 The second term in this equation corresponds to the statistical
	 fluctuation in the multiplication of electrons in the PMT.
	 Gauss(1) is a random value which follows a Gaussian distribution with mean
	 = 0 and sigma = 1. 
	 $C_{gain}$ is defined from the charge distribution 
	 of 1 p.e. light, which is measured in a bench test by using an LED, 
	 and it is defined channel by channel (typically it is $\sim$60\%).
	 
   \item Electronics\\
	 The number of photoelectrons is converted to ADC counts ($ADC_{raw}$)
	 by multiplying another conversion factor ($C_{conv2}$) with $N_{obs}$.
	 $C_{conv2}$ is measured from the 1 p.e. distribution obtained by LED light, and it is typically 
	 $\sim$ 60 ADC counts/p.e.

	 The non-linearity of electronics is simulated with an empirical function:
	 \begin{equation}
	  ADC_{obs} = ADC_{raw}/(1 + C_{nonlin} \times ADC_{raw})
	 \end{equation} 
	 where $C_{nonlin}$ is 0.000135/ADC counts. In case the ADC count is
	 greater than 4095, it is set to 4095 to account for saturation in the electronics.
  \end{enumerate}
  
\begin{figure}[h]
   \begin{center}
    \begin{minipage}{0.5\textwidth}
     \includegraphics[width=0.85\textwidth,clip]{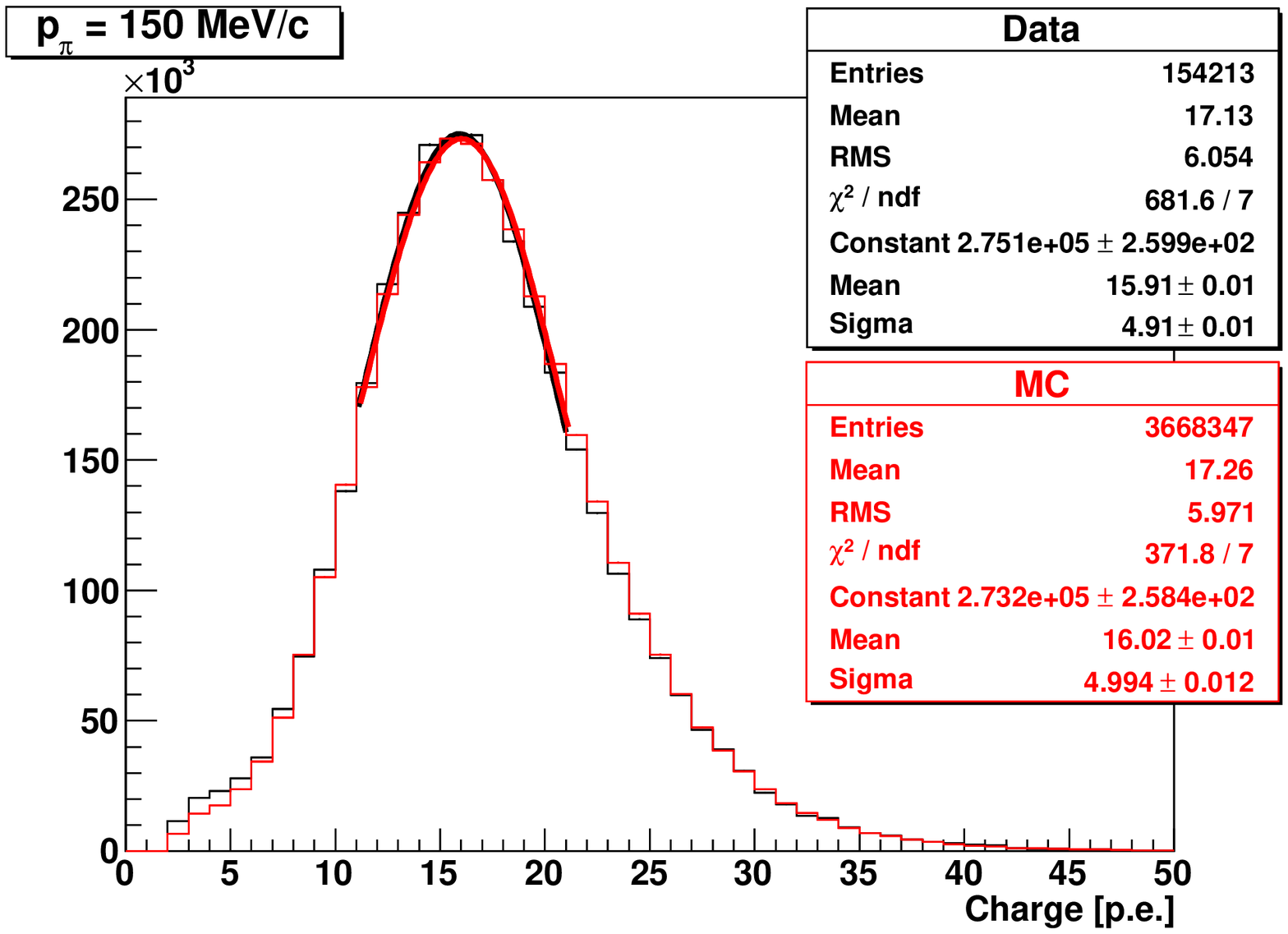}%
    \end{minipage}
    \hfill
    \begin{minipage}{0.5\textwidth}
     \includegraphics[width=0.85\textwidth,clip]{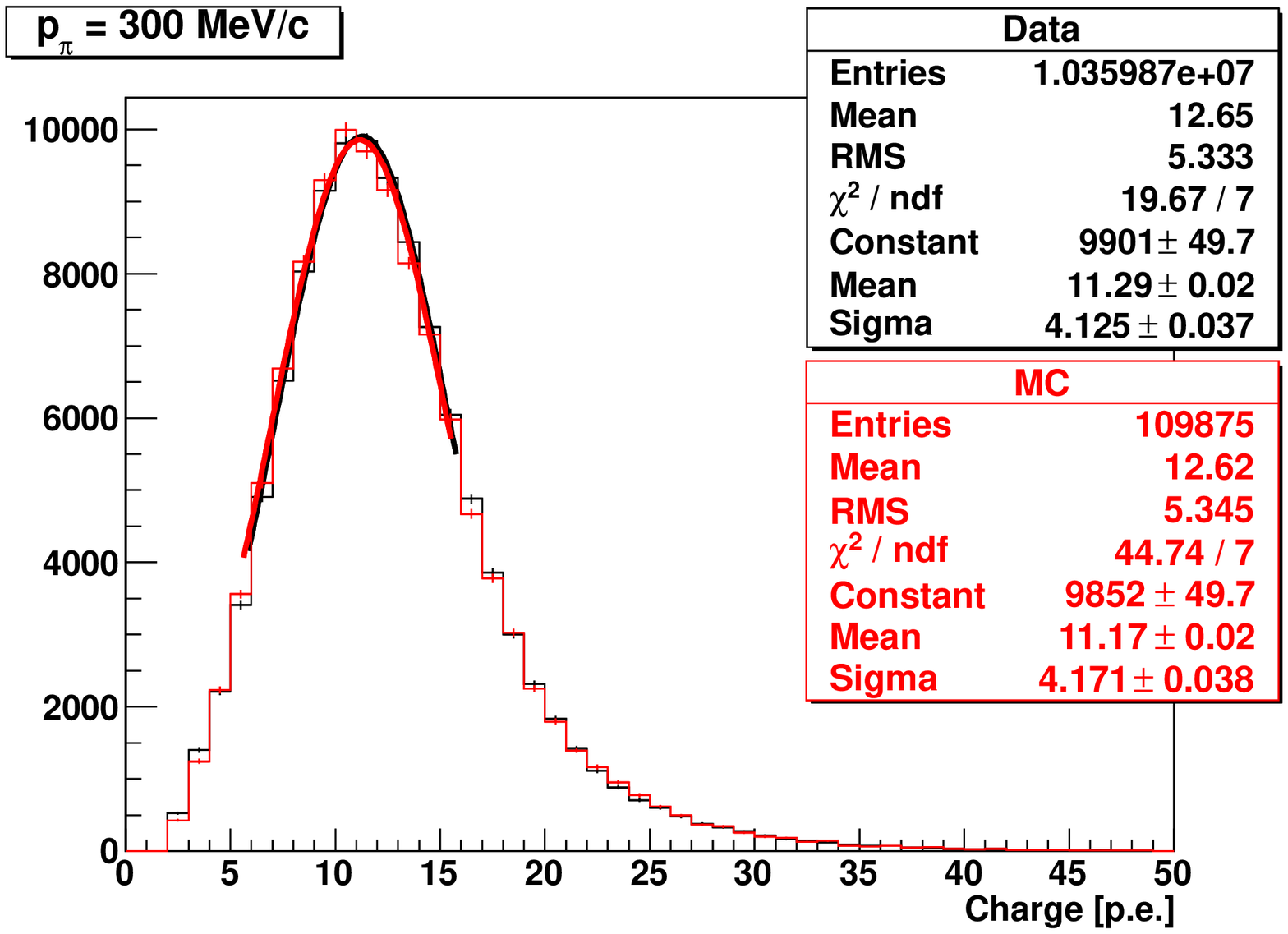}%
    \end{minipage}
    \caption{%
    Charge distribution of through-going pions for data and MC, for
    $p_{\pi}$ = 150 and 300 MeV/c data set. 
    The hits below 2.5 p.e. threshold are not shown in the plots.
    The fits closely follow each other.
    } \label{fig:charge_150_300}%
   \end{center}
  \end{figure}

  The conversion factor $C_{conv}$ and the non-linearity correction factor
  $C_{nonlin}$ are obtained by fitting the charge distributions of through-going
  pions with $p_\pi$ = 150 and 300 MeV/c.  Figure
  \ref{fig:charge_150_300} shows the charge distribution for data,
  compared with MC after the fit. The charge distribution in MC reproduces the
  distribution in data very well.

  The crosstalk hits are also implemented in the simulation. For each of
  the ``real'' hit associated with a particle trajectory,
  crosstalk hits are generated in adjacent channels in the MAPMTs. The expected
  number of photons for these crosstalk hits are calculated by
  multiplying the ``real'' hit by the crosstalk probability. The crosstalk
  probability in MC is tuned so that the charge distribution of
  crosstalk hits in the through-going pion data agree with data.
  In this tuning, crosstalk hits are selected from the hits which
  were not on the pion track. The crosstalk 
  probability for adjacent channels in a MAPMT is determined to be $\sim$2\%,
  and the crosstalk between adjacent fibers due to light leaking
  through the reflective coating is determined to be $\sim$0.8\%.

\begin{figure}[h]
   \begin{center}
    \includegraphics[width=0.45\textwidth]{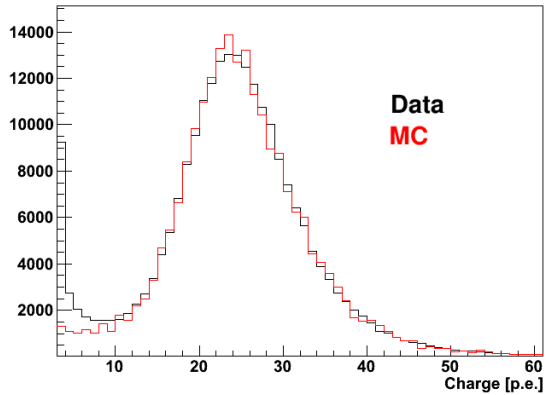}
   \end{center}
   \caption{
   CEMBALOS charge distribution of through-going muons for $p_{\pi} = $ 237.2 MeV/c setting.
   }
   \label{fig:charge_harp}
  \end{figure}

  The simulation and calibration procedure for the scintillating bars of CEMBALOS is the
  same as for the FGD. Figure \ref{fig:charge_harp} shows
  the charge distribution for through-going muons in CEMBALOS for the
  $p_\pi = $ 237.2 MeV/c setting, for data and MC (hereafter, the 237.2
  MeV/c data set will be used to show  an example). 
  The agreement between data and MC is good
  except for the low p.e. region. The disagreement in the low p.e. region is
  due to MPPC noise hits which is not implemented in the simulation. 
  Those noise hits are random and small (typically 1$\sim$3 p.e.). We
  apply a 5 p.e. threshold in the analysis to reject those hits.

\begin{figure}[h]
   \begin{center}
    \begin{minipage}{0.5\textwidth}
     \includegraphics[width=0.9\textwidth,clip]{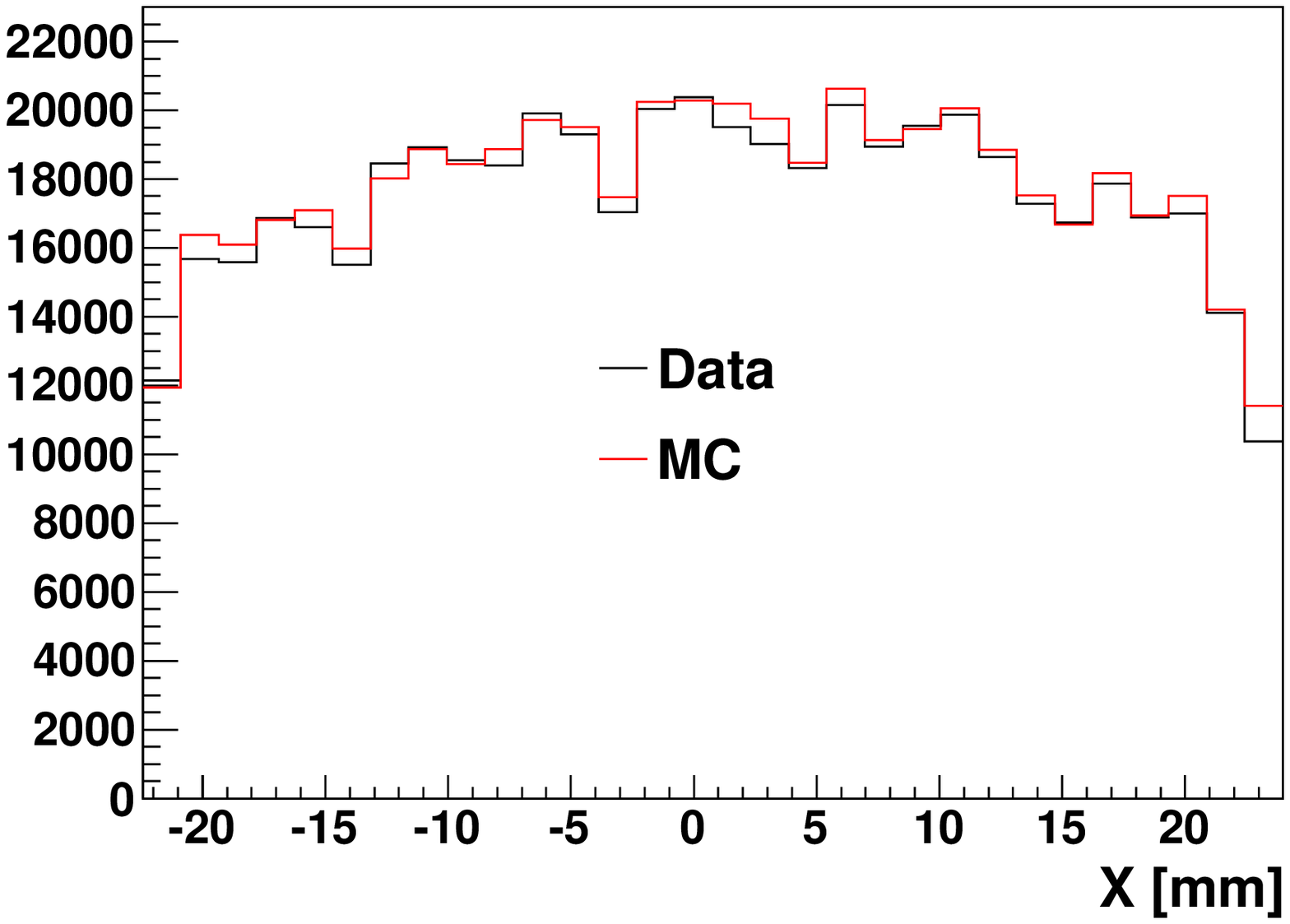}%
     \caption{%
     Beam position distribution in X, for the data set with $p_\pi =$ 237.2 MeV/c
     setting. The black (red) histogram shows the distribution for data (MC).
     } \label{fig:beam_pos}%
    \end{minipage}
    \hfill
    \begin{minipage}{0.5\textwidth}
     \includegraphics[width=0.9\textwidth,clip]{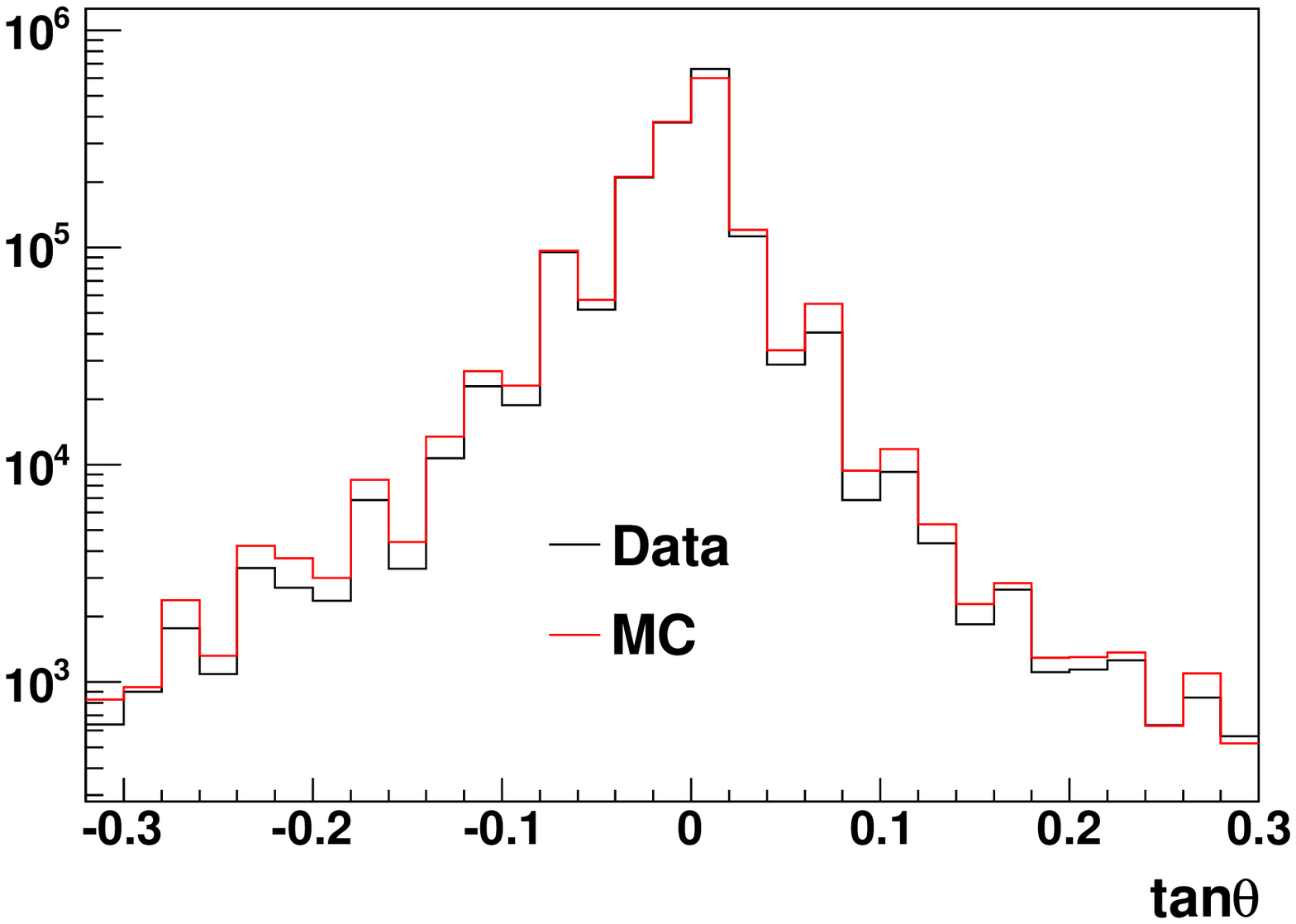}%
     \caption{%
     Beam angular distribution in X projection, for the data set with
     $p_\pi =$ 237.2 MeV/c setting. The variable $\theta$ is the angle
     from horizontal line (X=0).
     The black (red) histogram shows the distribution for data (MC).
     } \label{fig:beam_ang}%
    \end{minipage}
   \end{center}
  \end{figure}

  The beam position distribution and momentum are measured in data
  and reproduced in the simulation. In the simulation, pions are generated 1 cm upstream from the S0
  trigger. The X and Y position of the generation point and the angular distribution
  of the beam are tuned so that the measured beam position distribution
  and the angular distribution of the through-going tracks in the fiber
  tracker agree between data and MC. A Gaussian distribution is assumed
  for the initial position distribution and the angular distribution,
  and the mean and sigma of the distributions are tuned for X and
  Y. Figure \ref{fig:beam_pos} and \ref{fig:beam_ang} shows the beam 
position distribution and angular distribution for data with the 237.2
  MeV/c setting compared to the distribution for MC after tuning.

  \subsection{Data acquisition and event summary}
  The data acquisition is controlled by using MIDAS (Maximum
  Integration Data Acquisition System)\cite{midas}. 
  It controls the front-end DAQ programs for each detector,
  and combines the data to build events. 

  The data used in the analysis we describe in the following
  section is for a $\pi^+$ beam on a scintillator (carbon) target for five incident
  momenta (201.6, 216.6, 237.2, 265.5, 295.1 MeV/c) as has been discussed earlier. 
  There were $\sim$ 1.5 million beam triggered events recorded
  for each momentum settings, except for the 216.6 MeV/c setting where 
  0.5 Million events were recorded due to limited beam time. 

  \clearpage


  \section{\label{sec:selection}Event selection}

   \subsection{Event reconstruction} \label{sec:pi_reco}

\begin{figure}[hp]
    \includegraphics[width=7.5cm]{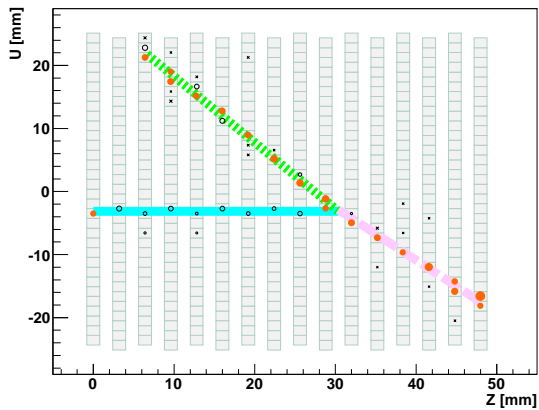}
    \caption{
    Example of ABS candidate event in data ($p_\pi=$237.2MeV/c).
    The filled circles (red) correspond to the large hits ($> 20$ p.e.), the
    crosses correspond to the hits identified as crosstalk hits and
    the thick lines (blue, green and red) correspond to reconstructed tracks.
    }
    \label{fig:abs_disp}
   \end{figure}

   As an illustration of the reconstruction, an ABS candidate event in 
   the data is shown in Figure \ref{fig:abs_disp} in the UZ projection
   where Z is the direction of the beam. The upstream horizontal (blue) track is identified as a pion 
   (``pion-like'' track). The other tracks (green and pink) are ``proton-like'' tracks
   produced by nuclei receiving energy from the incident $\pi^+$.
      
   We describe the track reconstruction procedure in the following section.
   
   The first step of the event reconstruction is the conversion from ADC
   count to the number of photo-electrons, followed by an electronics non-linearity correction.
   The typical number of p.e. is $\sim$ 11 p.e./hit for minimum ionizing
   particles, and only the hits above 2.5 p.e. are used in the track
   reconstruction. The efficiency to detect a hit  
   for charged particles passing through the layer is $\sim$93\%, where the inefficiency is caused by the 
   inactive region of the fiber. To minimize the effect of the inactive
   region the positions of the fiber layers are shifted relative to each
   other, as shown in Figure \ref{fig:abs_disp}.  
 
   In the track reconstruction algorithm, the candidate hits and
   crosstalk hits are treated differently. The crosstalk hits usually have smaller p.e. 
   and they are associated with hits with larger p.e. Hence when there is a 
   hit with a large p.e. ($>$ 20), the hits with smaller p.e. ($<$ 10)
   in the adjacent MAPMT channels are identified as crosstalk hits.
   The tracks are reconstructed in U and V layers individually, and then
   combined to make 3D tracks according to the following procedure.

   \begin{enumerate}
    \item Incident track search:\\
	  Track candidates are identified by 
	  searching for hits on straight trajectories.
	  For the incident track, the straight lines are required to
	  start from the upstream-most layer, and the angle of the lines are
	  required to be nearly horizontal (0 $\pm$ 4 degrees).
	  Starting from the hits in the upstream-most layer, hits on
	  the straight line are searched for in the downstream layers. 
	  Hits within two fiber-widths are included for the track candidate, 
	  with the process continuing towards subsequent layers until no
	  such hits are found. 
	  At least 3 hits are required to make
	  a track. 
	  The hits on the incident track are required to be not large ($<$
	  20 p.e.), so that the hits from a secondary proton track are not
	  included. In case the hits are large or identified as cross talk, it
	  is not used in the $>$ 3 hits requirement, but the hit
	  tracing does not stop. When there are multiple incident track
	  candidates, the longest track is selected.
    \item Interaction vertex search:\\
	  The end position of the incident track is selected as a temporary
	  interaction vertex point. Then a search is conducted for a best vertex
	  position around the temporary vertex in $\pm$ 3 layers and $\pm$
	  1 fiber region, where the best vertex position is defined as the
	  position where the largest number of hits can be traced.
	  The procedure to trace the hits is the same as that for the incident
	  track, except for the horizontal track requirement and small
	  hit ($<$ 20 p.e.) requirement. The tracks traced from the best vertex
	  position to the subsequent layers are selected as final tracks. 
    \item Combining the 2D tracks into a 3D track:\\
	  If the track ends of the 2D tracks in the U and V projections agree,
          the 2D tracks are combined to form a 3D track.
	  The track end positions may
	  not agree when the particle escapes the fiber crossing region
	  and leaves hits in only one projection. Otherwise the track
	  end position is required to agree within $\pm$ 2 layers. 
	  The Z position of the interaction vertex is defined as the
	  average Z position in two projections. The event is rejected
	  in the event selection if the Z position difference between
	  two projections are greater than 4.9 mm. 	  
   \end{enumerate}

\begin{figure*}[ht]
    \begin{center}
     \begin{minipage}{8.5cm}
      \includegraphics[width=7.3cm,clip]{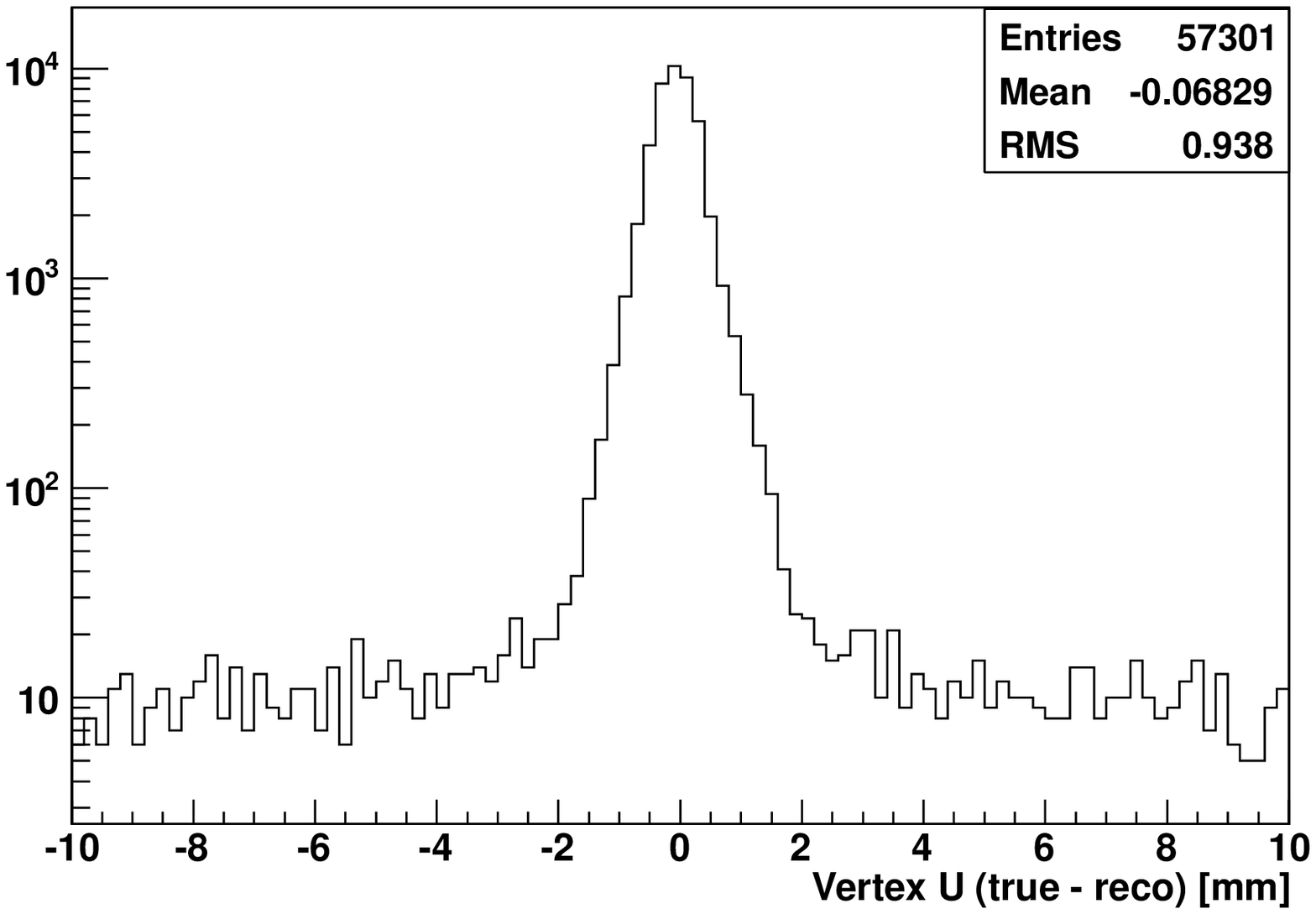}%
     \end{minipage}
     \begin{minipage}{8.5cm}
      \includegraphics[width=7.3cm,clip]{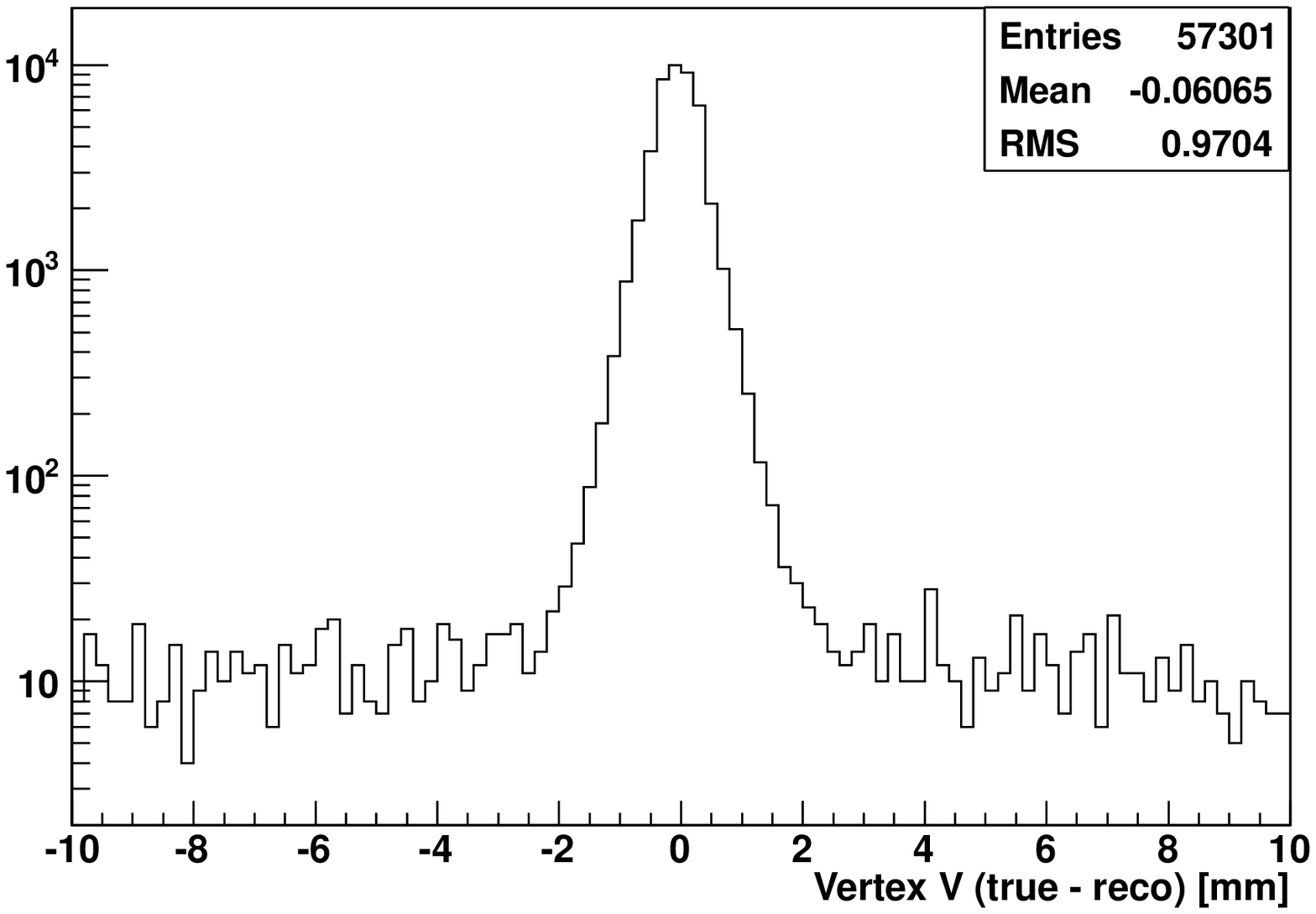}%
     \end{minipage}
     \begin{minipage}{8.5cm}
      \includegraphics[width=7.3cm,clip]{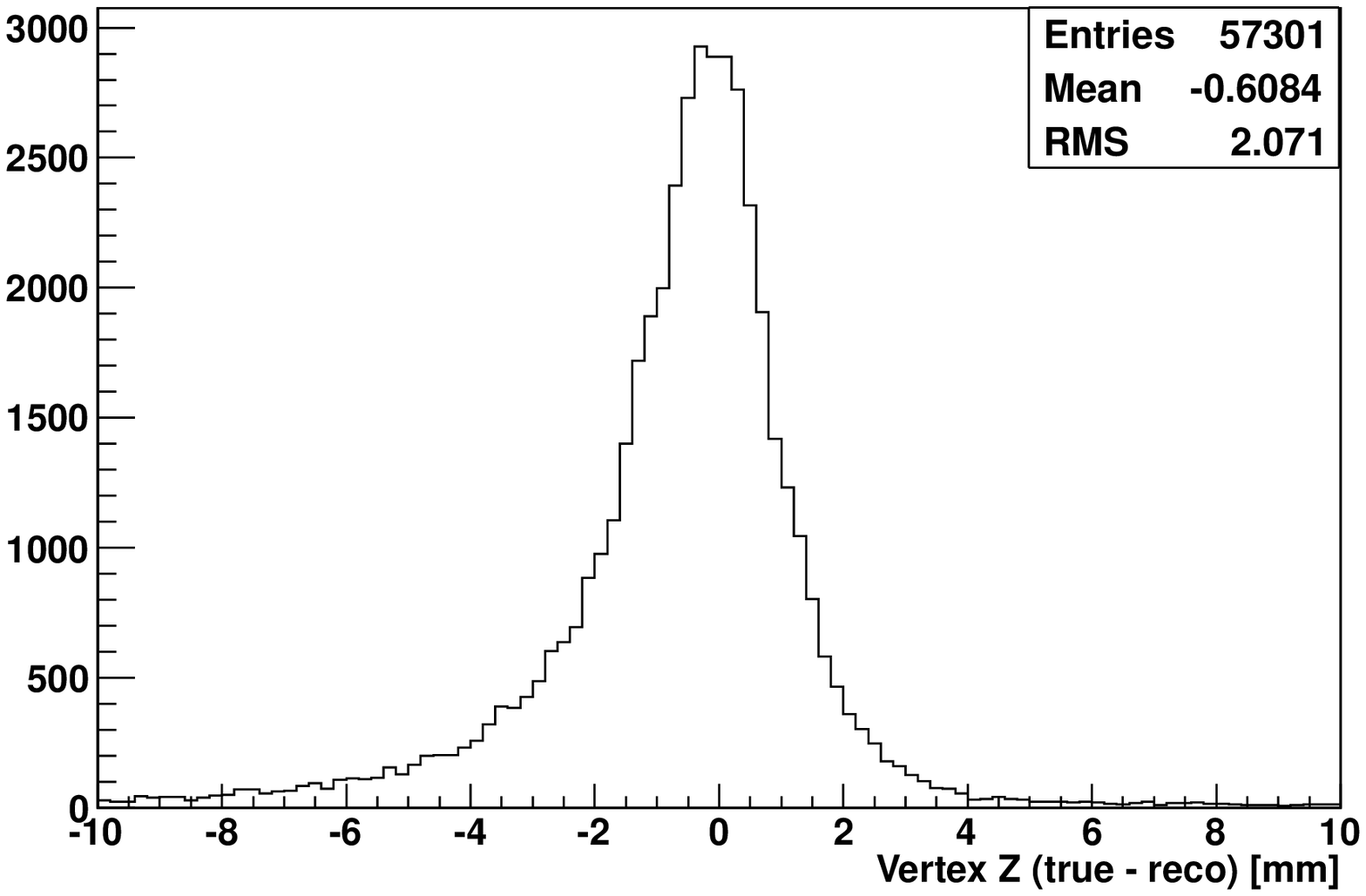}%
     \end{minipage}
     \begin{minipage}{8.5cm}
      \includegraphics[width=7.3cm,clip]{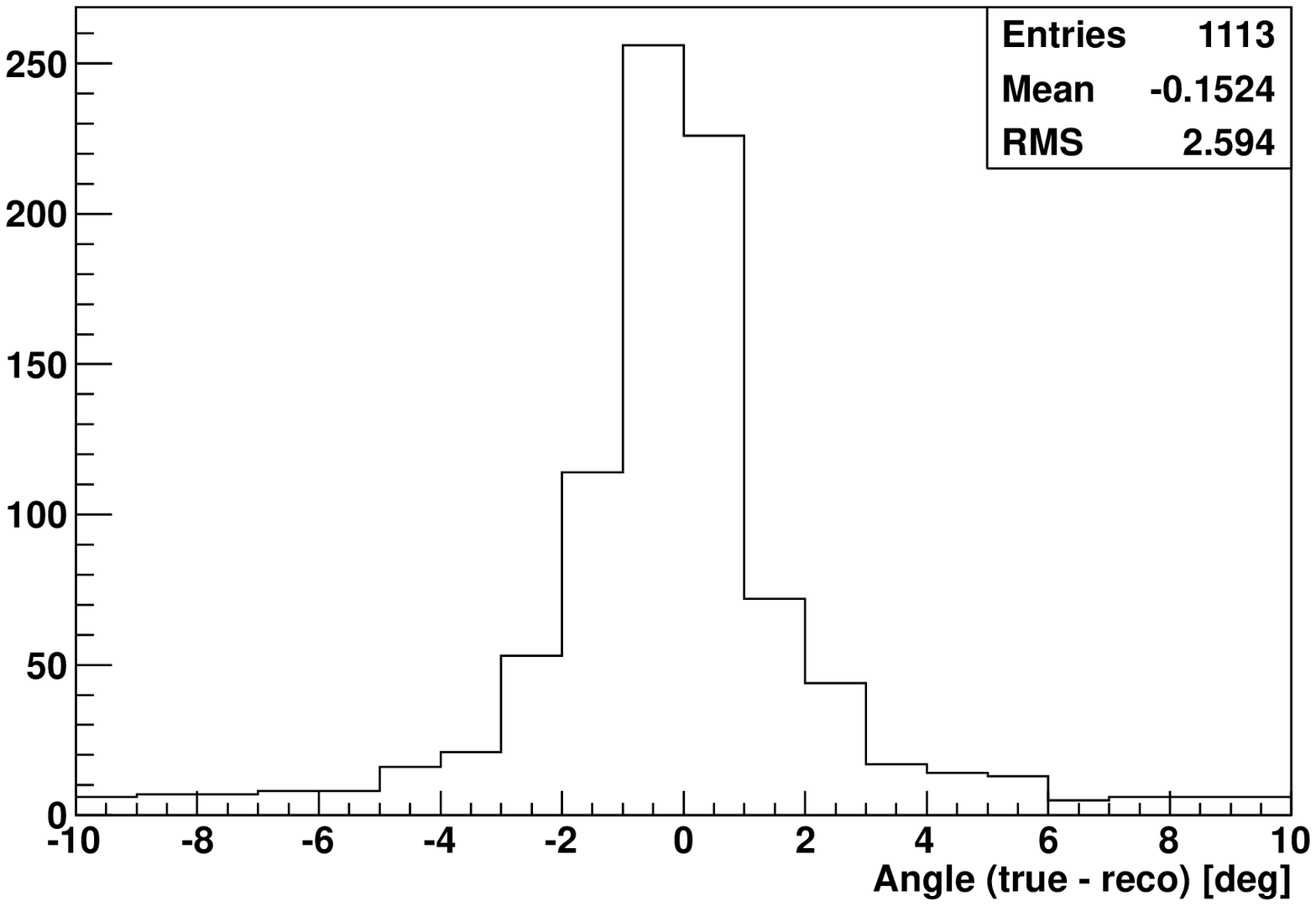}%
     \end{minipage}
     \caption{%
     Difference between the true and reconstructed  vertex position in U, 
     V and Z, and the true and reconstructed angle. 
     } \label{fig:vertex}%
    \end{center}
   \end{figure*}

   Comparing the reconstructed track with the true trajectory in the MC, the
   position resolution of the interaction vertex is evaluated to be $\sim$ 1 mm in U 
   and V, and $\sim$ 2 mm in Z (Figure \ref{fig:vertex} a,b,c). The angular resolution of the reconstructed
   track is evaluated to be $\sim$ 3 degree (Figure \ref{fig:vertex} d). 
   
   For each track, we calculate the deposited charge per track length,
   dQ/dx, obtained by dividing the total charge deposit by the
   total length of the track. The dQ/dx is used for identifying the particle types in the event selection. For large hits ($\sim$30 p.e.), the
   measured charge can be smaller than the actual charge 
   because of the electronics saturation effect. 
   The effect of saturation becomes significant when the path length within a fiber is long, 
   resulting in a large charge deposit. Since the angle relative the fiber orientation 
   in the U and V projections are different, the path length in each view will generally be different.
   In order to minimize the
   saturation effect, we calculate the dQ/dx from the
   projection with the shorter path length per fiber.

   \subsection{Event selection criteria} \label{sec:abscx_sel}


 \begin{figure}[hbt]
    \begin{center}
     \includegraphics[width=7.9cm]{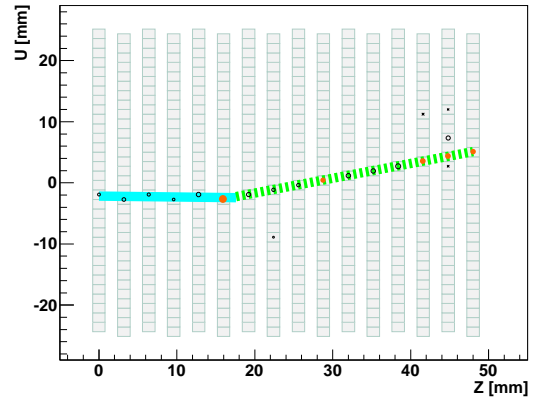}%
     \caption{%
     Example of pion scattering candidate event in data
     ($p_\pi=$237.2MeV/c). The track (blue) in the upstream side is identified as the incident pion
     track, and the track in the downstream side (green) is identified as a scattered pion track.
     } \label{fig:scat_disp}%
    \end{center}
   \end{figure}
   \begin{figure}[hbt]
    \begin{center}
     \includegraphics[width=7.8cm]{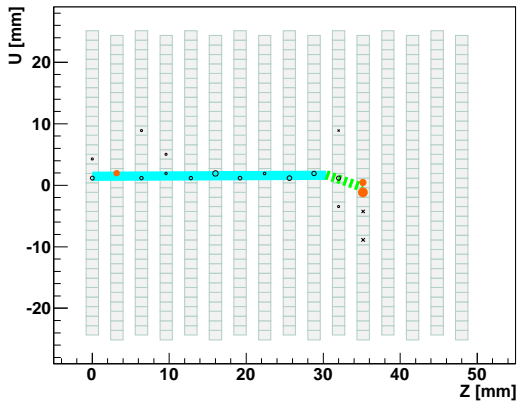}%
     \caption{%
     Example of CX candidate event in data ($p_\pi=$237.2MeV/c). The 
     track in the upstream side (blue) is assumed to be the incident
     pion track, and the track in the downstream side (green) is 
     assumed to be a proton track from CX interaction.
     } \label{fig:cx_disp}%
    \end{center}
   \end{figure}


  

   Examples of ABS, CX, SCAT event candidates are shown in Figure \ref{fig:abs_disp}, 
   \ref{fig:cx_disp}, and \ref{fig:scat_disp} respectively. The SCAT events can be readily 
   identified by the outgoing pion track, in contrast to the ABS and CX events where the 
   incident pion track terminates and may lead to the emission of proton tracks. CX events 
   are identified by a coincident signaling in the NaI crystals resulting from the outgoing photons 
   from the decay of the $\pi^0$  from the charge exchange reaction. As a result, ABS+CX events 
   are selected by requiring no $\pi^+$ in the final state, where final state tracks are identified 
   as all reconstructed tracks in the event apart from the incident pion, whereas SCAT events 
   have a pion track in the final state in addition to any protons that may be  produced 
   in the interaction. ABS events typically have one or two protons in the final state, whereas
   a CX event will usually have zero or only one proton.

   The ABS + CX event selection is considered in further detail below.

\begin{figure}[htbp]
    \begin{center}
     \includegraphics[width=8.0cm]{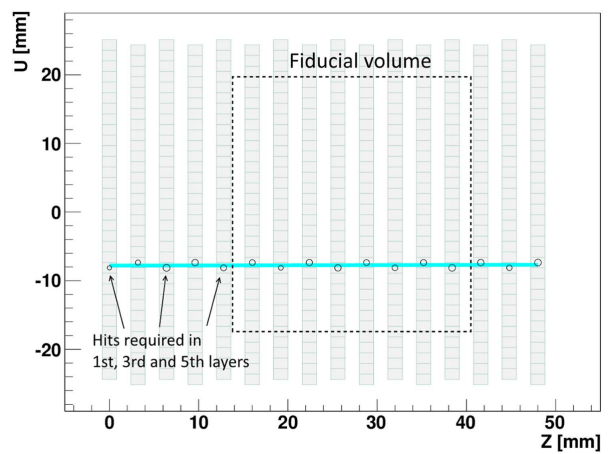}%
     \caption{%
     Illustration of the {\it Good incident $\pi^+$} cut requirement. The broken
     line represents the boundary of the fiducial volume
     } \label{fig:goodinc}%
    \end{center}
   \end{figure}

\begin{figure}[htbp]
    \begin{center}
     \includegraphics[width=8.0cm]{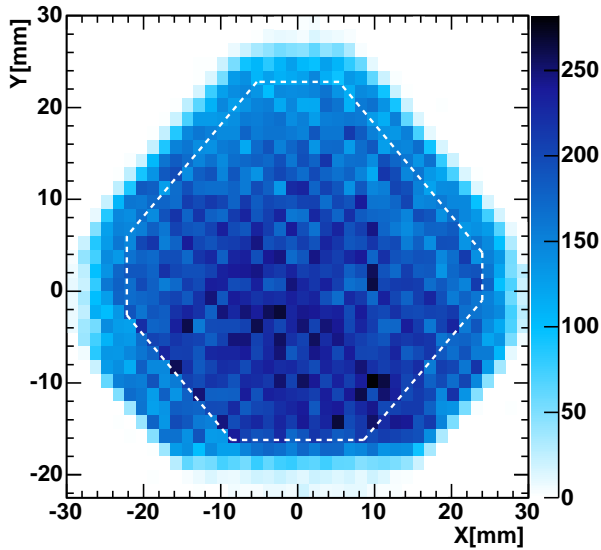}%
     \caption{%
     The X-Y view of incident beam position distribution. The white broken
     line represents the boundary of the fiducial volume
     } \label{fig:beam_2d}%
    \end{center}
   \end{figure}

\begin{figure*}[htbp]
    \begin{center}
     \begin{minipage}{5.9cm}
      \includegraphics[width=6cm,clip]{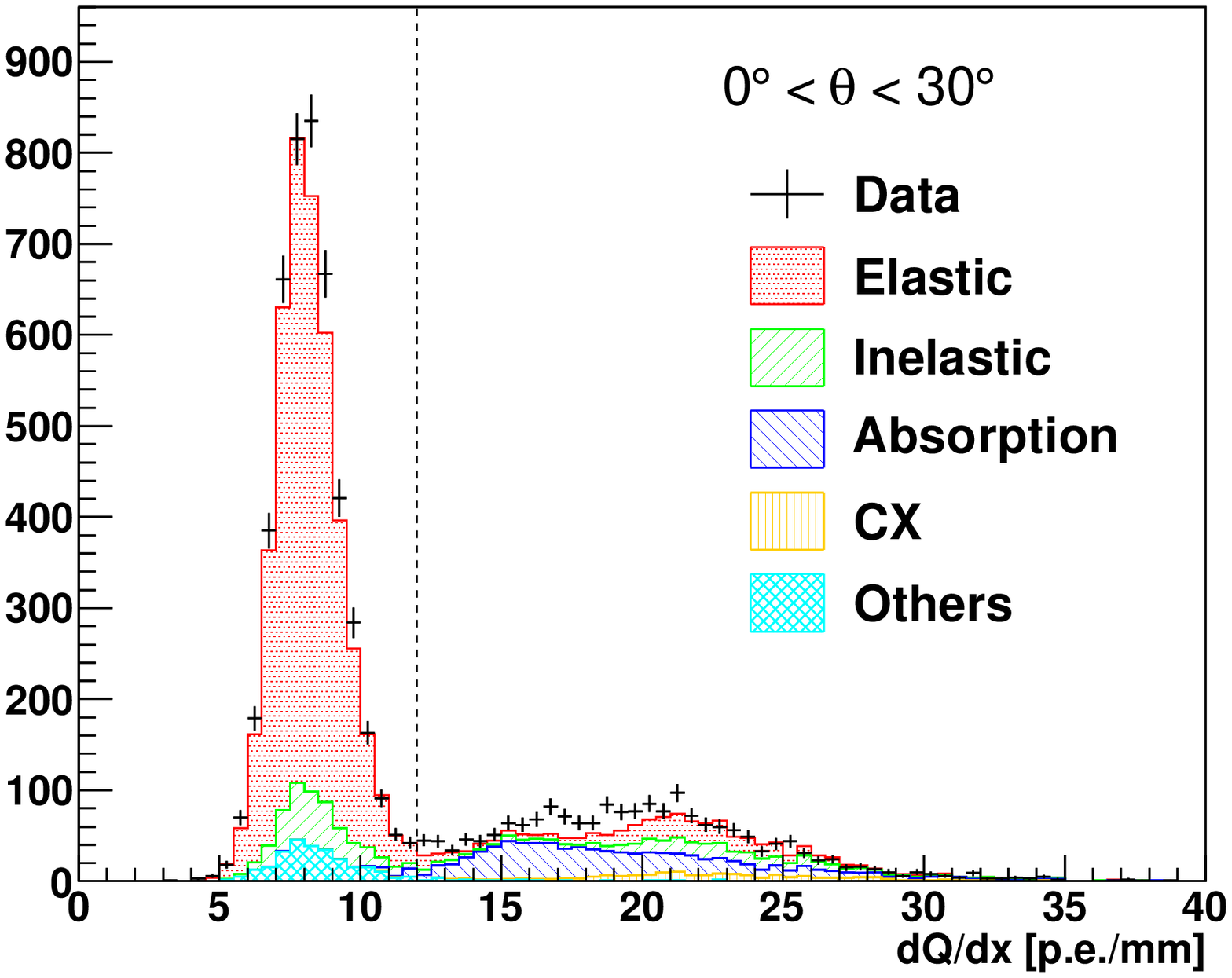}%
     \end{minipage}
     \hfill
     \begin{minipage}{5.9cm}
      \includegraphics[width=6cm,clip]{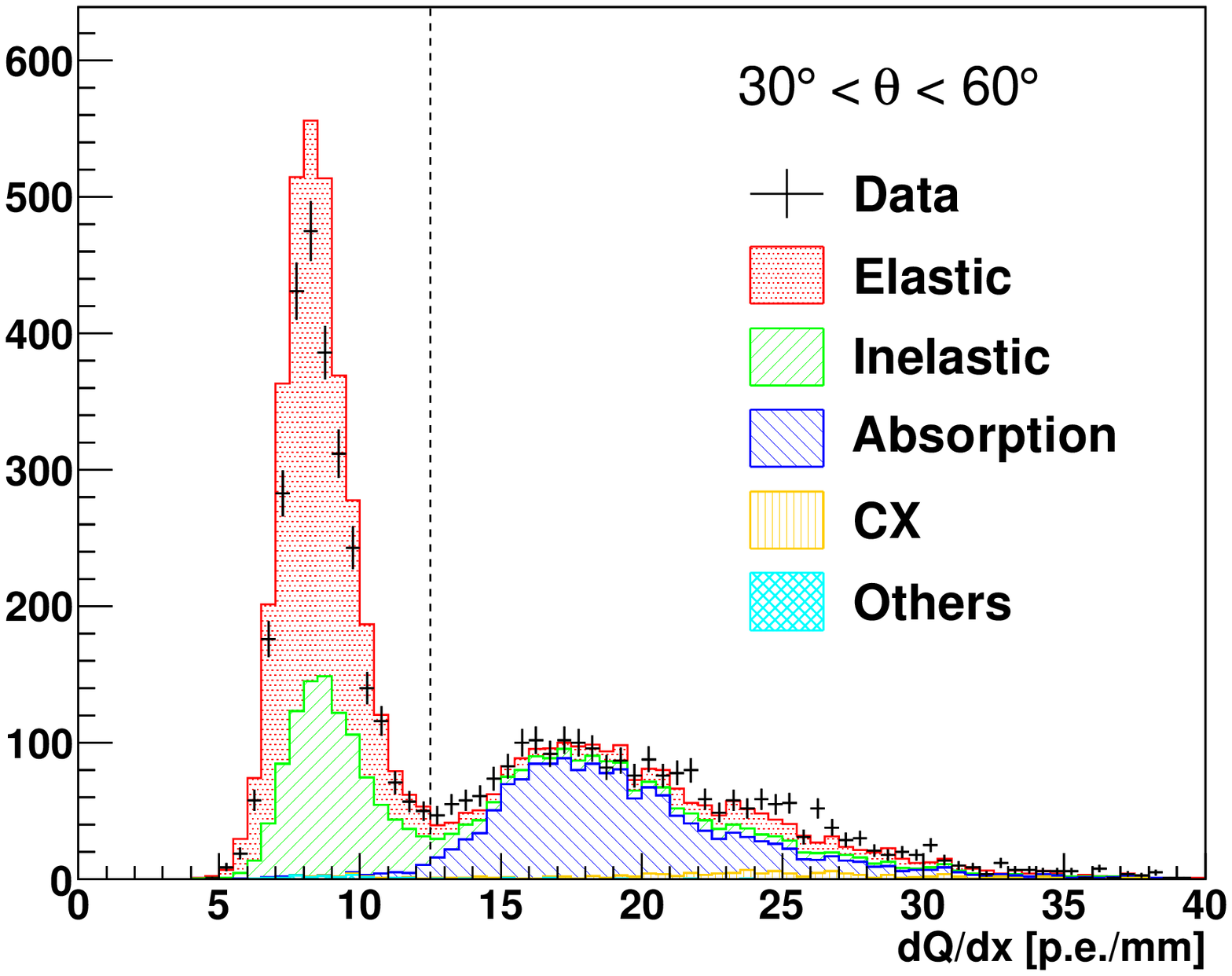}%
     \end{minipage}
     \hfill
     \begin{minipage}{5.9cm}
      \includegraphics[width=6cm,clip]{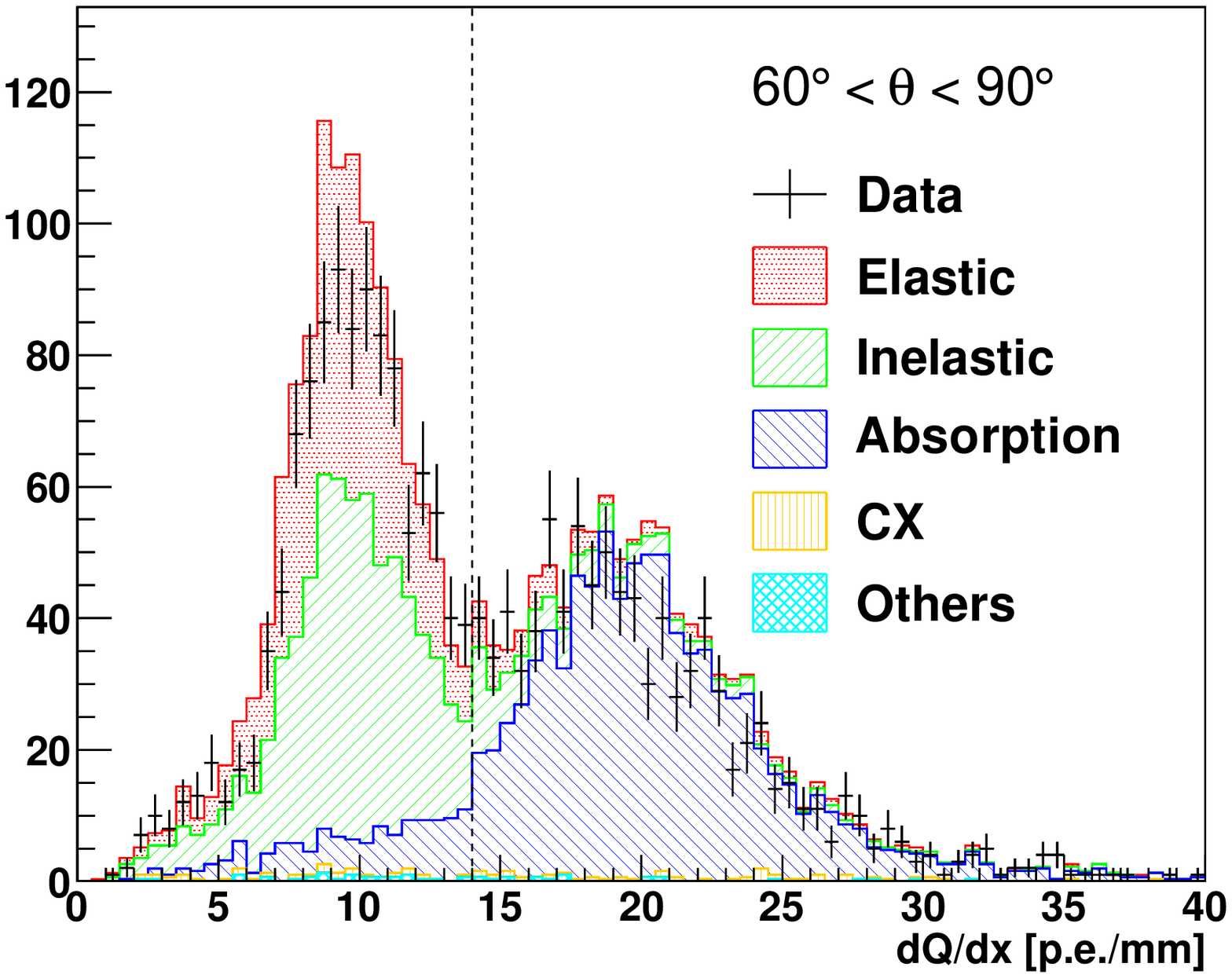}%
     \end{minipage}
     \begin{minipage}{5.9cm}
      \includegraphics[width=6cm,clip]{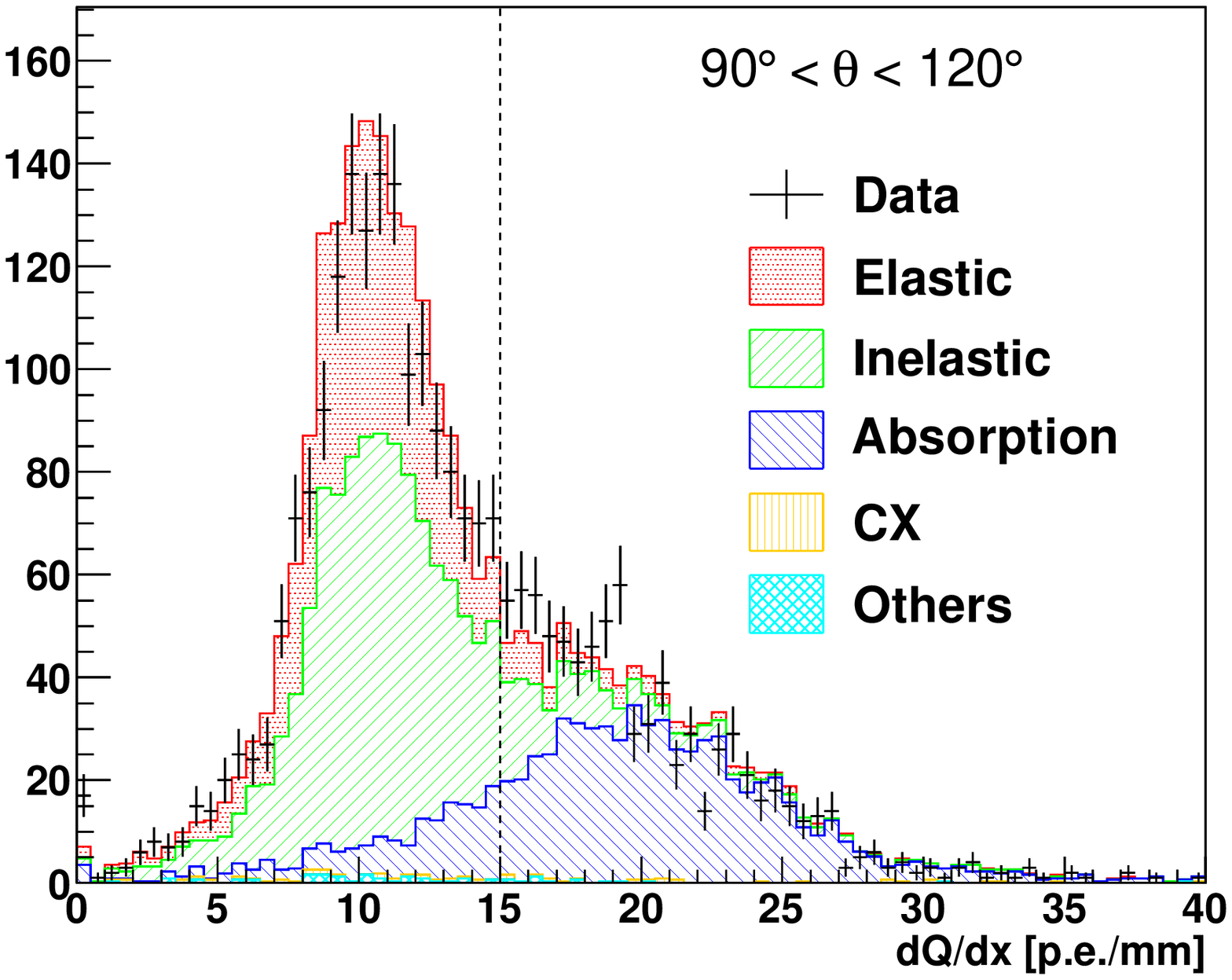}%
     \end{minipage}
     \hfill
     \begin{minipage}{5.9cm}
      \includegraphics[width=6cm,clip]{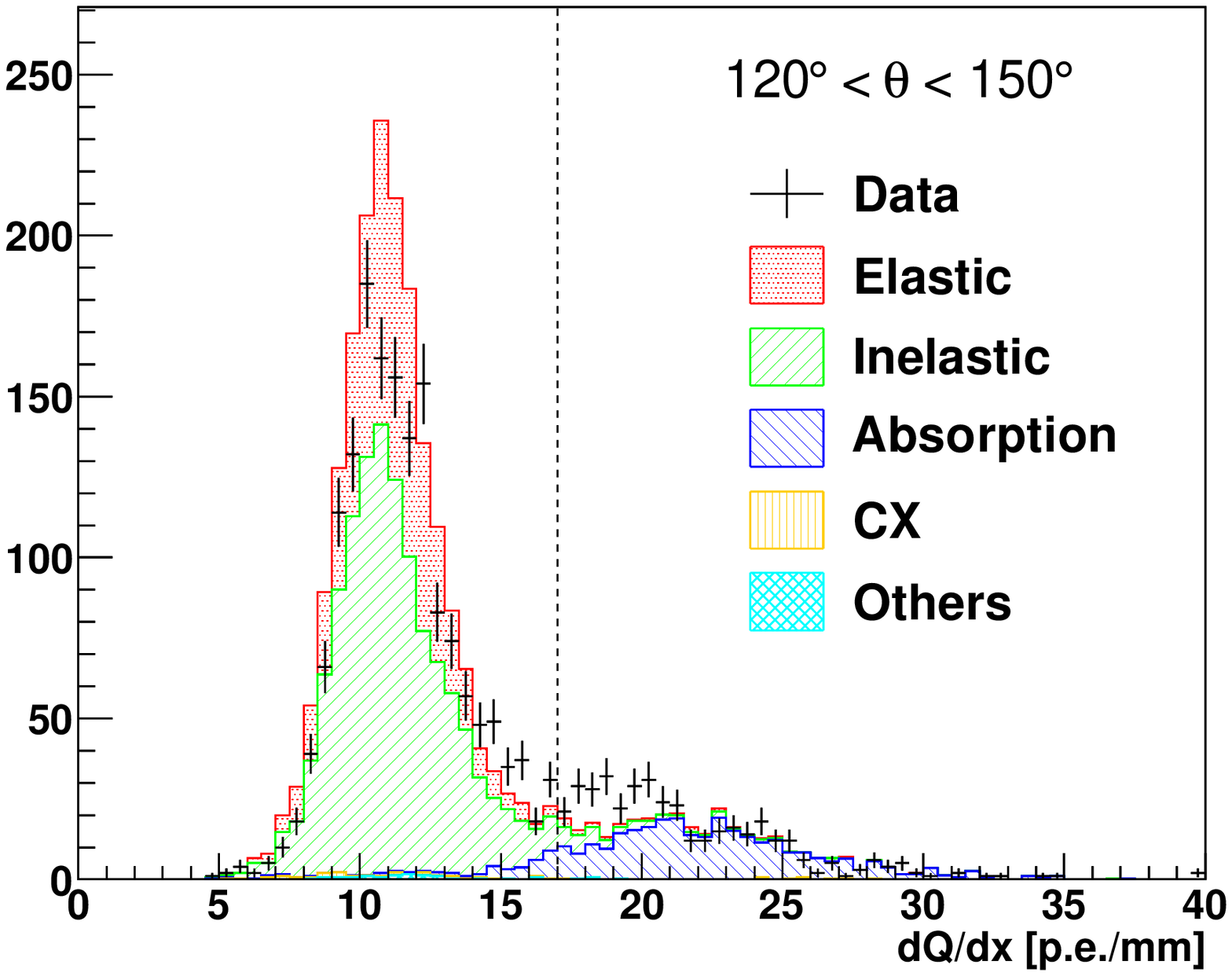}%
     \end{minipage}
     \hfill
     \begin{minipage}{5.9cm}
      \includegraphics[width=6cm,clip]{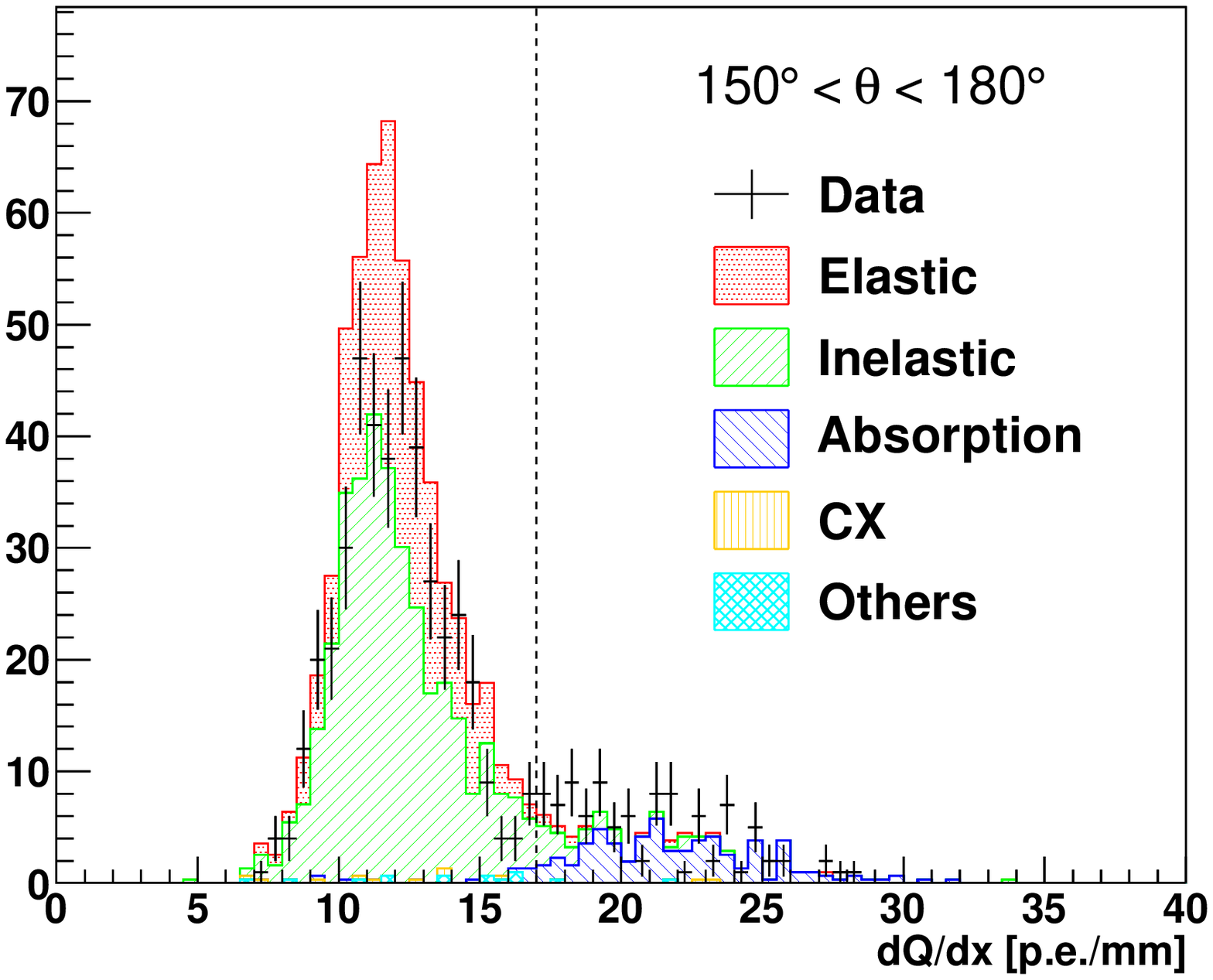}%
     \end{minipage}
     \caption{%
     dQ/dx distribution in six different angular regions for $p_\pi=$ 237.2 MeV/c for data
     and MC. The dotted vertical lines represent the threshold to
     distinguish pions (left of the line) and protons. For multiple track events, only the
     smallest value of dQ/dx among the tracks is filled in the
     histogram. The events in the ``Others'' category is mainly from events with pions decaying 
     in flight and Coulomb scattering events.
     } \label{fig:dqdx_250}%
    \end{center}
   \end{figure*}

   \begin{enumerate}
    \item {\it Good incident $\pi^+$}\\
	  This selection consists of three requirements.
	  First, we require that the incident particle is a charged 
	  pion. We apply a cut in the Cherenkov light vs. TOF distribution,
	  as explained in Section \ref{sec:beamline} (except for the 201.6
	  MeV/c data set, in which we used the TOF distribution only).

	  Second, we impose requirements to make sure that a straight track,
	  normal to the incidence plane, exists. For this we require hits on the first, third and fifth layers,
	  and in the same fiber position (i.e. same U, V position) in both the U and V projections (see Figure \ref{fig:goodinc}).
	  Only a horizontal straight track passes this cut. The
	  background muons originating from the decay of pions are 
	  rejected by this cut because in most of cases the angle of
	  these muons are shifted with respect to the beam axis. 
	  
	  Third, we require the incident track to enter the fiducial volume (FV).
	  The FV is shown as the broken lines in Figure \ref{fig:goodinc}
	  and \ref{fig:beam_2d}. Figure \ref{fig:beam_2d} shows the X, Y
	  position distribution of the incident beam. The hexagonal shape
	  corresponds to the region where the S1 trigger overlaps with the
	  fiber crossing region. Because the reconstruction algorithm
	  requires at least 3 hits to reconstruct a track, the
	  fiducial volume is defined to be $\ge$ 3 fibers (3
	  layers) from the upstream edge of the fiber crossing
	  region. The X, Y
	  position of the incident track is required to be inside the X-Y
	  plane of the FV.
    \item {\it Vertex in the FV} \\
	  After the {\it Good incident $\pi^+$} selection, $\sim$90\% of the remaining events
	  are through-going pion events. The events with pion interactions
	  are selected by requiring a reconstructed vertex inside the FV.
	  With this cut we attempt to reject not only through-going events 
	  but also pion scattering events with a 
	  very small scattering angle (``small angle'' event). 
	  To identify these events, we count the number of
	  hits inside or outside $\pm$ 2 fibers of the incident U, V
	  position. ``Small angle'' events look very similar to
	  through-going pion events, but can be rejected by
	  requiring no reconstructed hits outside the 2 fiber region and 
	  $\ge$25 hits inside the 2 fiber region, with at
	  least 2 hits in the last three layers.  
    \item {\it No final $\pi^+$}  track\\
	  In this selection we require there be no $\pi^+$ in the final state.
	  The pion tracks are distinguished from proton tracks by applying a 
	  dQ/dx cut. Figure \ref{fig:dqdx_250} shows an example of dQ/dx
	  distributions for $p_\pi = $ 237.2 MeV/c for data and MC. There are
	  six plots corresponding to six different angular regions 
	  ($0^{\circ} < \theta < 30^{\circ}, 30^{\circ} < \theta < 60^{\circ},... 
	  150^{\circ} < \theta < 180^{\circ}$), where $\theta$ is the angle of the
	  reconstructed track with respect to the beam direction.
	  The histograms for MC are normalized by the number of incident
	  pions. The color of the histograms represents the interaction
	  types. The vertical broken line
	  represents the threshold to distinguish pions and
	  protons. Because the dQ/dx distribution varies with angle and incident momentum, 
	  different thresholds are set for each combination of outgoing track angle and incident momentum.
	  If any of the reconstructed tracks except the incident track
	  is found to have dQ/dx below the threshold, then that track is
	  identified as a charged pion, and the event is not
	  selected. \\
	  In order to identify the scattered pion track which is
	  reconstructed only in U or V projection, the dQ/dx cut is also
	  applied for the 2D tracks. 
	  For the 2D tracks, the dQ/dx is calculated by using
	  the track length projected in 2D, which is shorter than the
	  actual 3D track length. Therefore, the dQ/dx is overestimated
	  for 2D tracks. However, we apply the same dQ/dx threshold for both 3D and 2D
	  tracks, to avoid mis-identifying ABS or CX events as pion scattering events.
   \end{enumerate}


   \subsection{Selection efficiencies}
   The number of selected events after each stage of the cuts is summarized in Table
   \ref{tbl:event_reduc}. There are $\sim$7000 events in data after the
   event selection, except for the 216.6 MeV/c data set in which the number of
   incident pions is smaller due to the limited data taking time.
   The efficiency to select ABS or CX events which
   occurs inside the fiducial volume is estimated to be $\sim$79\%, and
   the purity of ABS + CX events in the selected sample is estimated to be
   $\sim$73\%. The details of the MC simulation and comparison with data
   after event selection are explained in Sec IV. 
   
   \begin{table*}[hp]
    \begin{tabular}{lcccccccccc}
     \noalign{\hrule height 1pt}
     & \multicolumn{2}{c}{201.6 MeV/c} & \multicolumn{2}{c}{216.6 MeV/c} &
     \multicolumn{2}{c}{237.2 MeV/c} & 
     \multicolumn{2}{c}{265.5 MeV/c} & \multicolumn{2}{c}{295.1 MeV/c}\\
     \cline{2-11}
     Cut & Data & MC & Data & MC & Data & MC & Data & MC & Data & MC \\
     \noalign{\hrule height 0.5pt}
     {\it Good incident $\pi^+$} & \multicolumn{2}{c}{273625} &
     \multicolumn{2}{c}{67164} & \multicolumn{2}{c}{276671} &
     \multicolumn{2}{c}{238534} & \multicolumn{2}{c}{282611} \\
     {\it Vertex in FV}        & 17522 & 18895.9 & 4833 & 5118.8 & 21861 & 22932.1 & 20567 & 20895.1 & 24327 & 24136.7 \\
     {\it No final $\pi^+$}          & 6797  & 6331.2  & 1814 & 1695.9 & 7671  &  7619.0 &  6772 & 7005.1  & 7289  & 7491.1 \\
     \noalign{\hrule height 0.5pt}
     Efficiency [\%] & \multicolumn{2}{c}{79.0} & \multicolumn{2}{c}{79.6} & \multicolumn{2}{c}{79.9} & \multicolumn{2}{c}{79.2} & \multicolumn{2}{c}{77.1} \\
     Purity [\%]    & \multicolumn{2}{c}{73.0} & \multicolumn{2}{c}{73.3} & \multicolumn{2}{c}{73.1} & \multicolumn{2}{c}{73.5} & \multicolumn{2}{c}{73.1} \\
     \noalign{\hrule height 1pt}
	\end{tabular}
    \caption{
    The number of events after each stage of the cut. The numbers for
    MC are normalized by the numbers of good incident pion events in data.
    }
    \label{tbl:event_reduc}
   \end{table*}

  \subsection{Background}
   When pions are scattered, the scattered pion tracks are not
	always well reconstructed, particularly when the pion
	is scattered nearly 90 degrees and the track passes between
	fiber layers. Also, due to finite dQ/dx resolution, pion tracks
	are sometimes misidentified as protons. These background events
	pass the event selection. Although the cross section of pion
	elastic scattering in the MC is tuned to results from previous experiments, a linear
	interpolation of the data points from the previous measurements does
	not perfectly reproduce the actual cross section.
	The estimation of the uncertainty for the number of predicted background
	events is described in section \ref{sec:bg_estimate}.

 \section{Simulation and tuning of physics models}
 \label{sec:phy_model}
  The hadronic interaction of the pions with a nuclei is simulated by using the
  list of physics models called ``QGSP-BERT''.  For the elastic scattering, it
  uses a model called ``hElasticLHEP'' based on a simple parametrization
  of the cross section. The inelastic scattering (INEL), ABS and CX are
  included in the inelastic process, which are simulated using the Bertini
  Cascade model\cite{bertini}.There are also other processes, namely double 
  charge exchange and hadron production, but the cross
  sections for those interactions are negligibly small in the pion
  momentum range in this experiment.
  \begin{figure}[ht]
   \begin{center}
    \includegraphics[width=8cm]{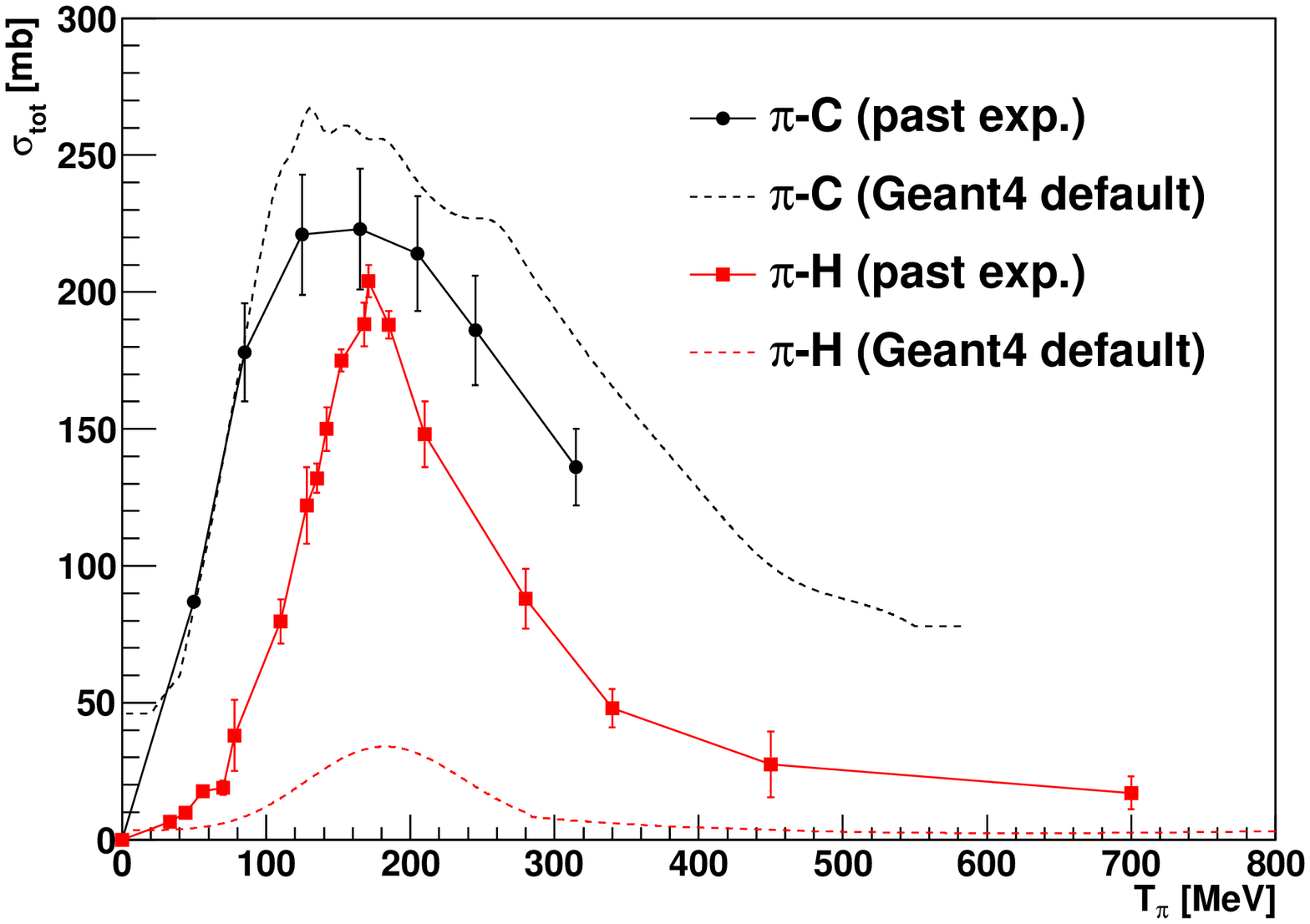}%
    \caption{%
    Comparison of elastic inclusive cross section between the previous
    experiments (summarized in Table \ref{tbl:xsec_past}) and the default Geant4. 
The cross sections are plotted as a function of pion
    kinetic energy.
    } \label{fig:xsec_elas}%
   \end{center}
  \end{figure}
  \begin{figure}[ht]
   \begin{center}
    \includegraphics[width=8cm]{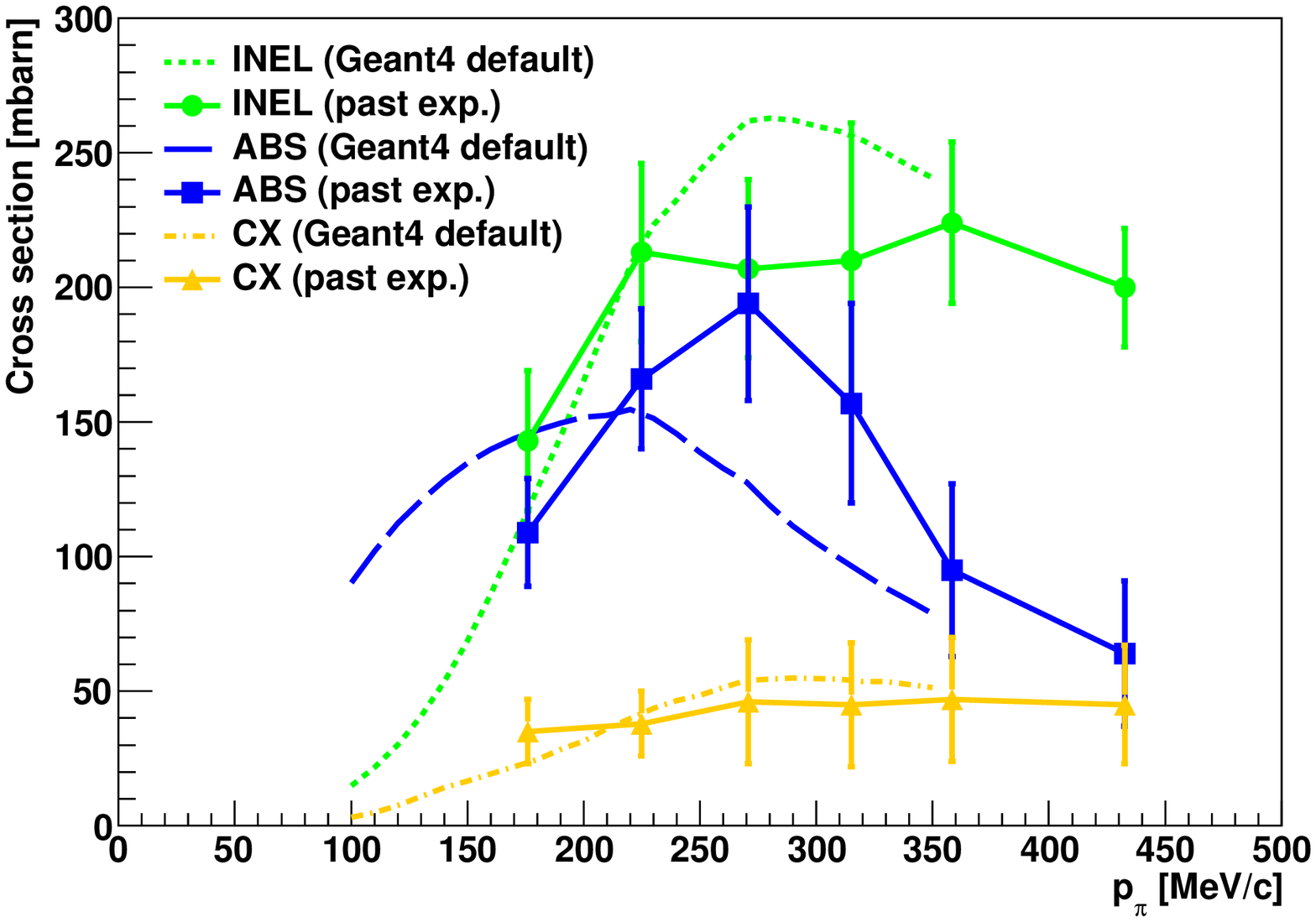}%
    \caption{%
    Comparison of inelastic inclusive cross sections between the previous
    experiment \cite{Ashery} and the default Geant4. The cross sections are
    plotted as a function of pion momentum.
    } \label{fig:xsec_inel}%
   \end{center}
  \end{figure}
  
  The $\pi^+-$C and $\pi^+-$H elastic cross sections and differential cross sections
  (d$\sigma/d\theta$) were tuned by interpolating the data points from previous measurements. 
  The inclusive $\pi^+-$C inelastic
  scattering, ABS and CX cross sections were also tuned. Figure
  \ref{fig:xsec_elas} and \ref{fig:xsec_inel} shows the comparison of
  the cross sections between the previous experiments and the default Geant4 MC data,
  for elastic and inelastic processes. There are disagreements between
  Geant4 cross section (ver9.4, QGSP-BERT) and the measurements
  from the previous experiments, especially for $\pi-$H elastic scattering
  process. 
  Table
  \ref{tbl:xsec_past} summarizes the data from previous experiments that we
  used for the tuning. The momentum of pions after inelastic scattering is predicted 
  using the NEUT cascade model \cite{NEUT} because there is no available data. 

  \begin{table*}[h]
   \begin{tabular}{lll}
    \noalign{\hrule height 1pt}
    Measurement & Kinetic energy (MeV) & Reference \\
    \noalign{\hrule height 0.5pt}
    $\pi-$C inclusive & \multirow{2}{*}{85, 125, 165, 205, 245, 315} & \multirow{2}{*}{D. Ashery et al. \cite{Ashery}} \\ 
    (elastic, inelastic, ABS and CX) & & \\
    $\pi-$C elastic inclusive & 49.9 & M. A. Moinester et al. \cite{moinester} \\
    \noalign{\hrule height 0.5pt}
    \multirow{5}{*}{$\pi-$H elastic inclusive} & 33, 44, 56, 70 & S. L. Leonard et al. \cite{leonard} \\
    & 78, 110, 135   & H. L. Anderson et al. \cite{anderson} \\
    & 165            & H. L. Anderson et al. \cite{anderson2} \\
    & 128, 142, 152, 171, 185 & J. Ashkin et al. \cite{ashkin} \\
    & 210, 280, 340, 450, 700 & Lindenbaum et al. \cite{lindenbaum} \\
    \noalign{\hrule height 0.5pt}
    \multirow{8}{*}{$\pi-$C elastic differential} & 40 & M. Blecher et al. \cite{Blecher} \\
    & 50 & R. R. Johnson et al. \cite{johnson} \\
    & 67.5 & J. F. Amann et al. \cite{amann} \\
    & 80 & M. Blecher et al. \cite{blecher2} \\
    & 100 & L. E. Antonuk et al. \cite{Antonuk} \\
    & 142 & A. T. Oyer et al. \cite{Alden} \\
    & 162 & M. J. Devereux et al. \cite{Michael} \\
    & 180, 200, 230, 260, 280 & F. Binon et al. \cite{Binon} \\
    \noalign{\hrule height 0.5pt}
    \multirow{5}{*}{$\pi-$H elastic differential} & 29.4, 49.5, 69 & J. S. Frank et al. \cite{Frank} \\
    & 69 & Ch. Joram et al. \cite{Joram} \\
    & 87, 98, 117, 126, 139 & J. T. Brack et al. \cite{Brack1} \\
    & 87, 98, 117, 126, 139 & J. T. Brack et al. \cite{Brack2} \\
    & 166.0, 194.3, 214.6, 236.3, 263.7, 291.4 & P. J. Bussey et al. \cite{Bussey} \\
    \noalign{\hrule height 1pt}
   \end{tabular}
   \caption{List of data sets used for cross section tuning in simulation.}
   \label{tbl:xsec_past}
  \end{table*}

\begin{figure}[h]
   \begin{center}
    \includegraphics[width=8cm]{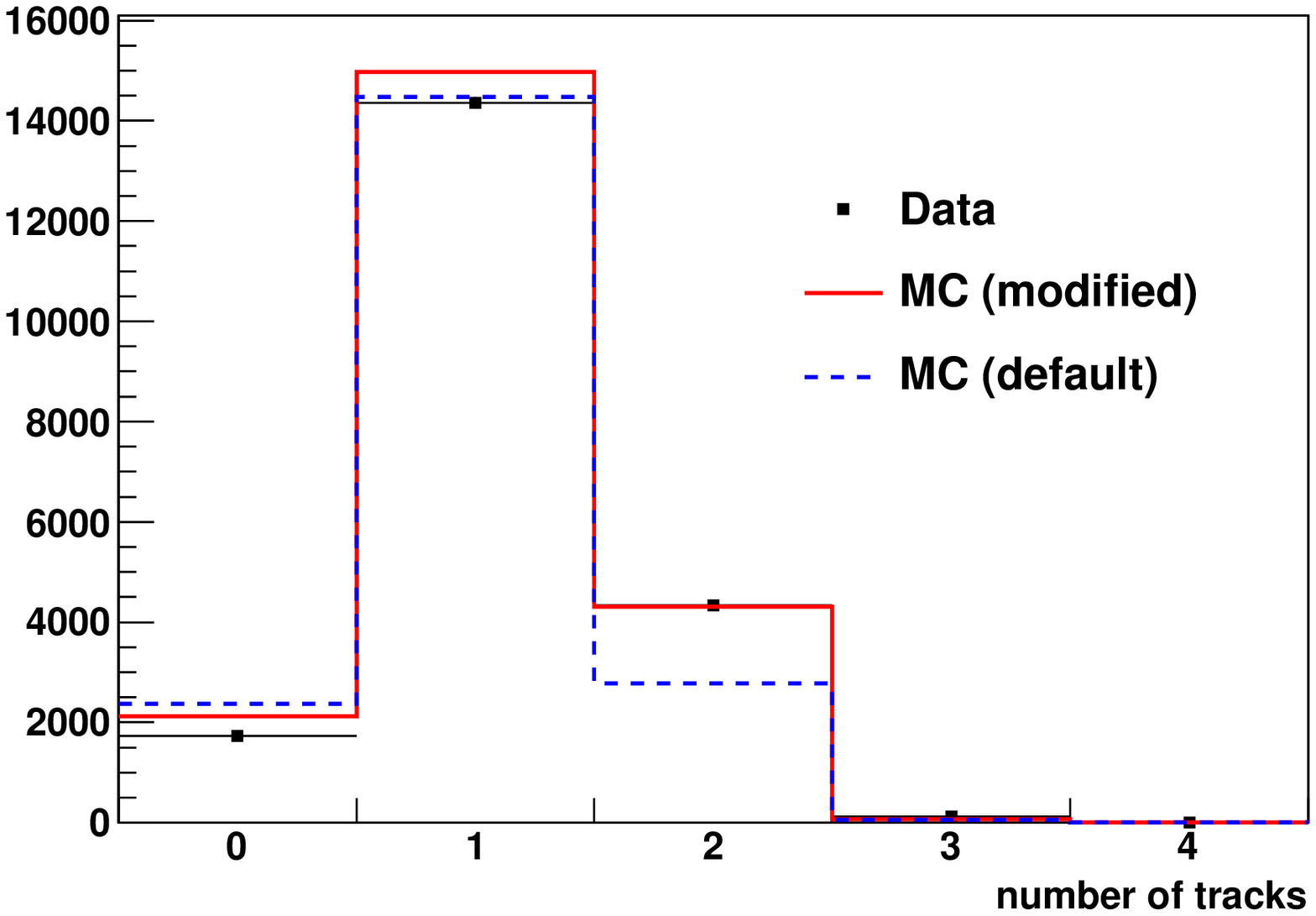}%
    \caption{%
    The number of reconstructed tracks for data, and the MC before and after tuning, for $p_\pi = $237.2
    MeV/c setting. The {\it No final $\pi^+$} cut is not applied.
    } \label{fig:tune_comp_ntracks}%
   \end{center}
  \end{figure}

\begin{figure}[h]
   \begin{center}
    \includegraphics[width=8cm]{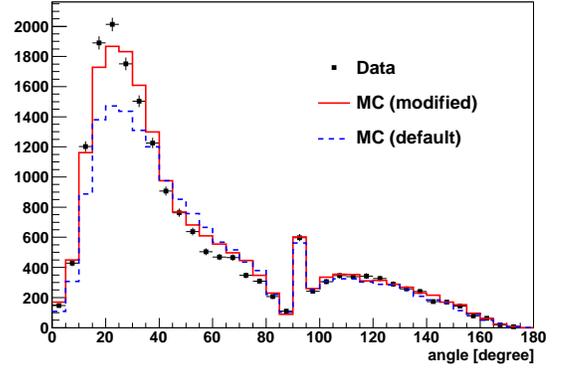}%
    \caption{%
    The angular distribution of reconstructed tracks
    for MC before and after tuning, and for data, for $p_\pi = $237.2
    MeV/c data set. The {\it No final $\pi^+$} cut is not applied.
    } \label{fig:tune_comp_angle}%
   \end{center}
  \end{figure}

  Figure \ref{fig:tune_comp_ntracks} and \ref{fig:tune_comp_angle} shows the number of
  tracks and angular distribution for the reconstructed tracks before
  and after the tuning, for $p_\pi$ = 237.2 MeV/c data set. The {\it No final $\pi^+$} cut
  is not applied for these plots. The forward angle multiple track
  events increased after the tuning, mainly due to the increase of $\pi-$H
  elastic cross section. The agreement between data and MC is much
  better with the tuning, although there are still small disagreements
  because the linear interpolation does not perfectly reproduce the
  data. The difference between data and MC is included in the systematic
  error. 

\begin{figure*}[htbp]
   \begin{center}
    \begin{minipage}{5.9cm}
     \includegraphics[width=6.0cm,clip]{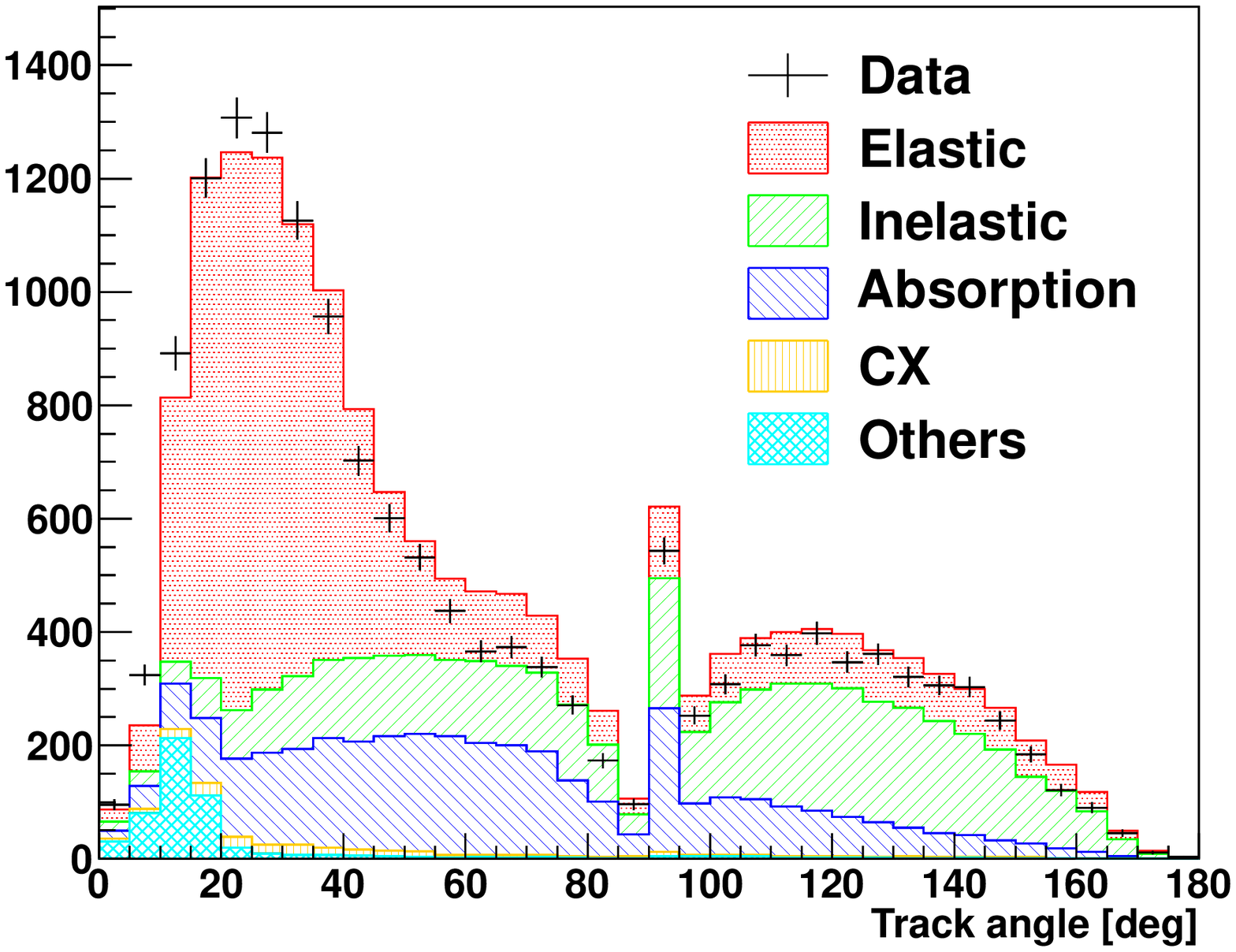}%
    \end{minipage}
    \begin{minipage}{5.9cm}
     \includegraphics[width=6.0cm,clip]{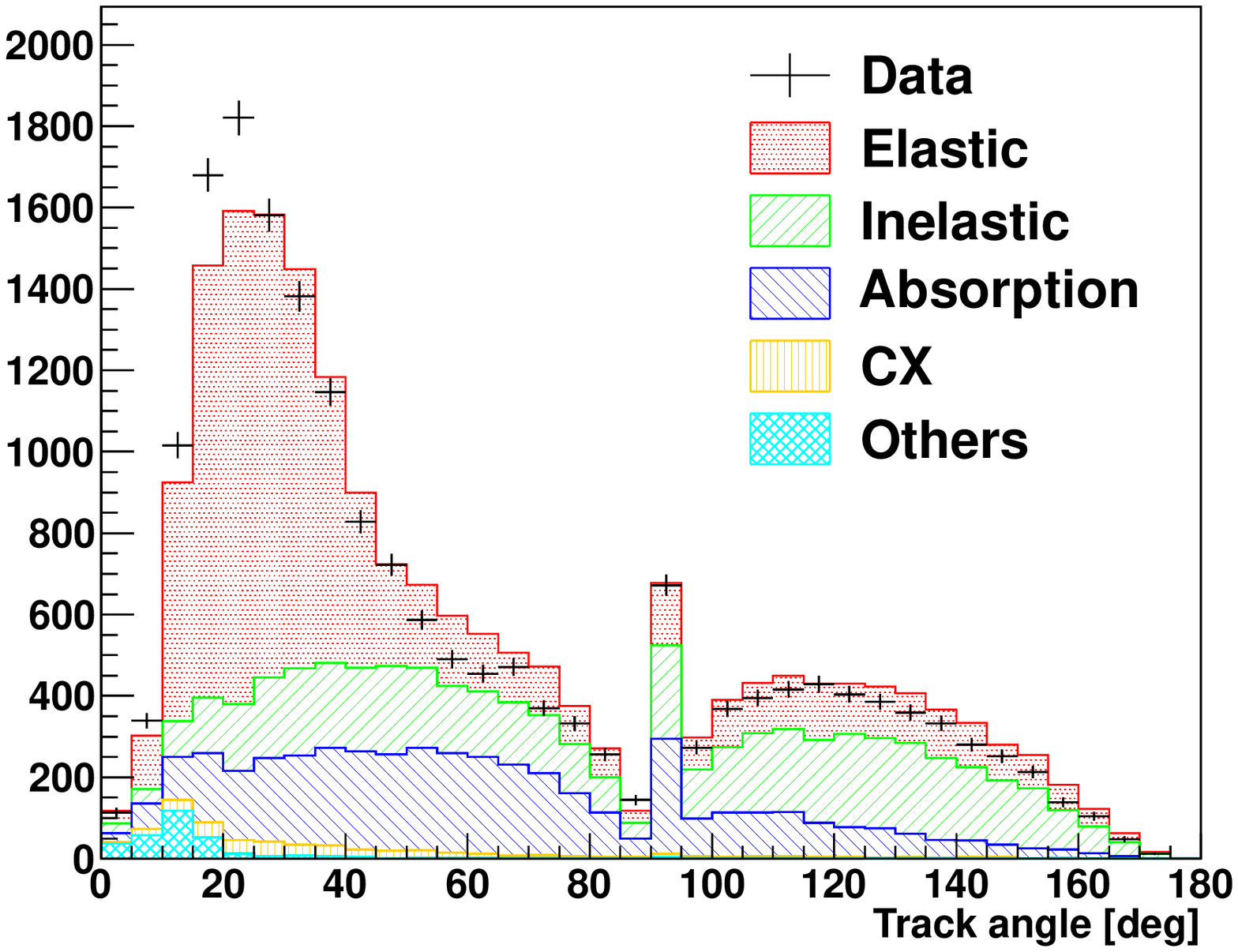}%
    \end{minipage}
    \begin{minipage}{5.9cm}
     \includegraphics[width=6.0cm,clip]{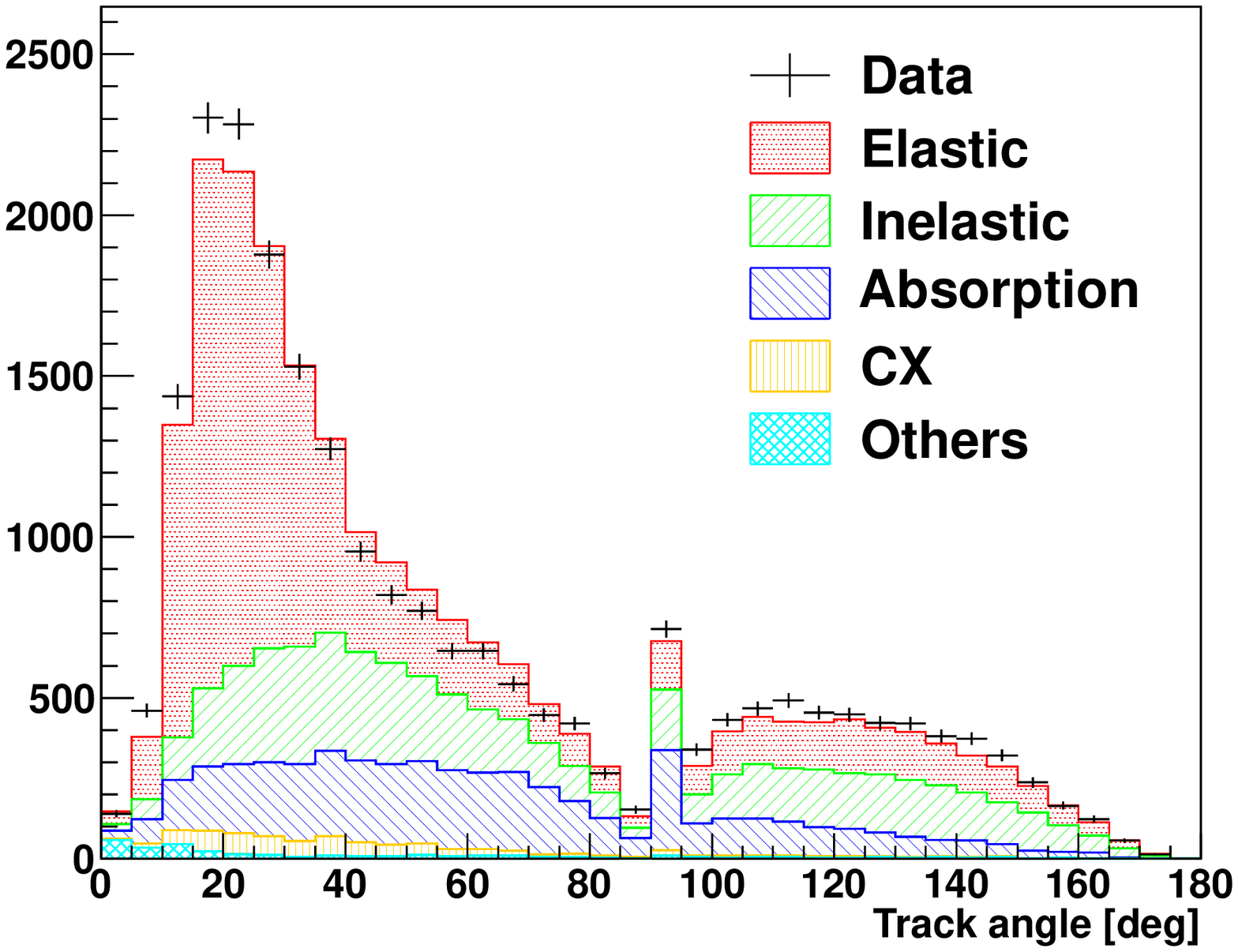}%
    \end{minipage}
    \begin{minipage}{5.9cm}
     \includegraphics[width=6.0cm,clip]{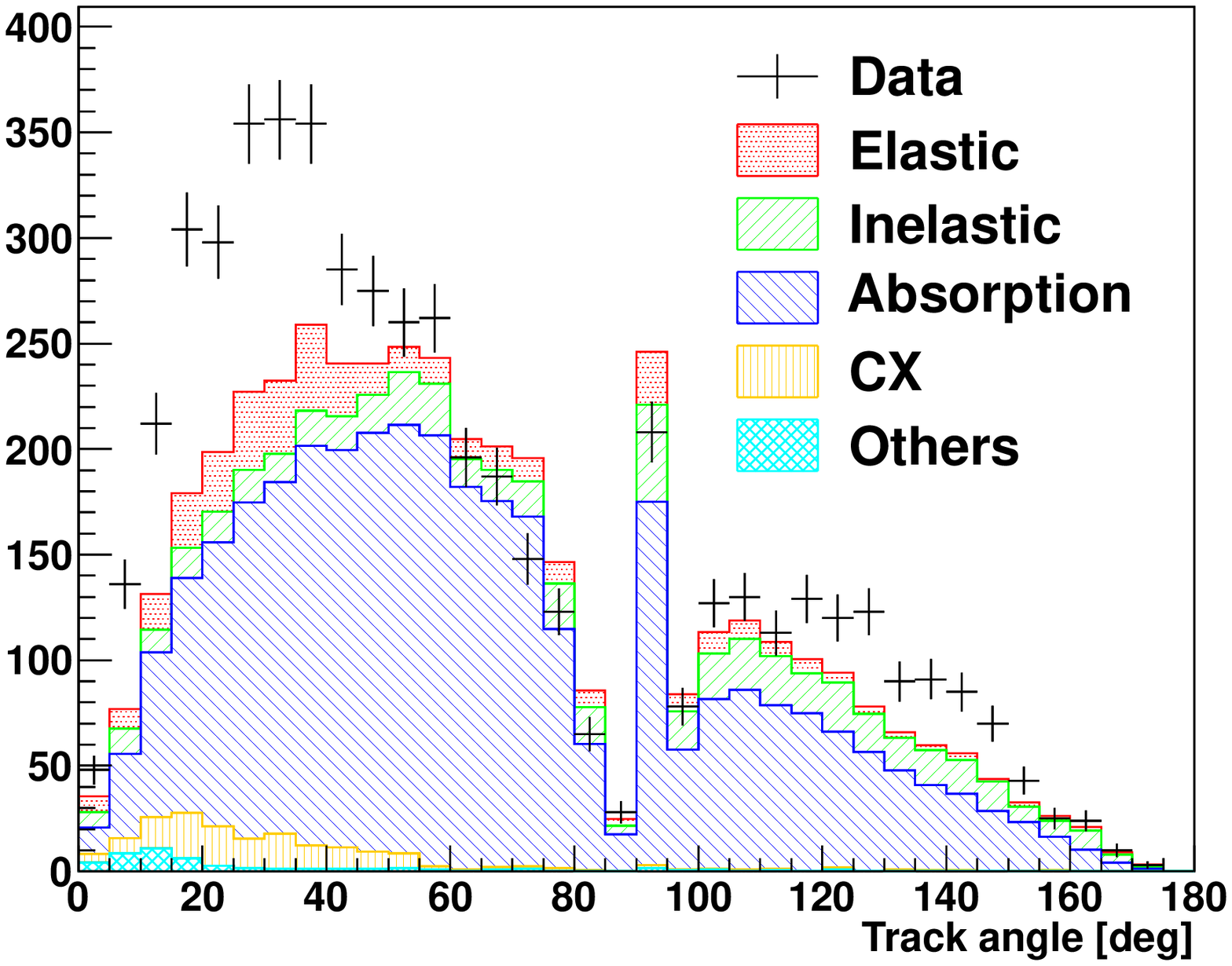}%
    \end{minipage}
    \begin{minipage}{5.9cm}
     \includegraphics[width=6.0cm,clip]{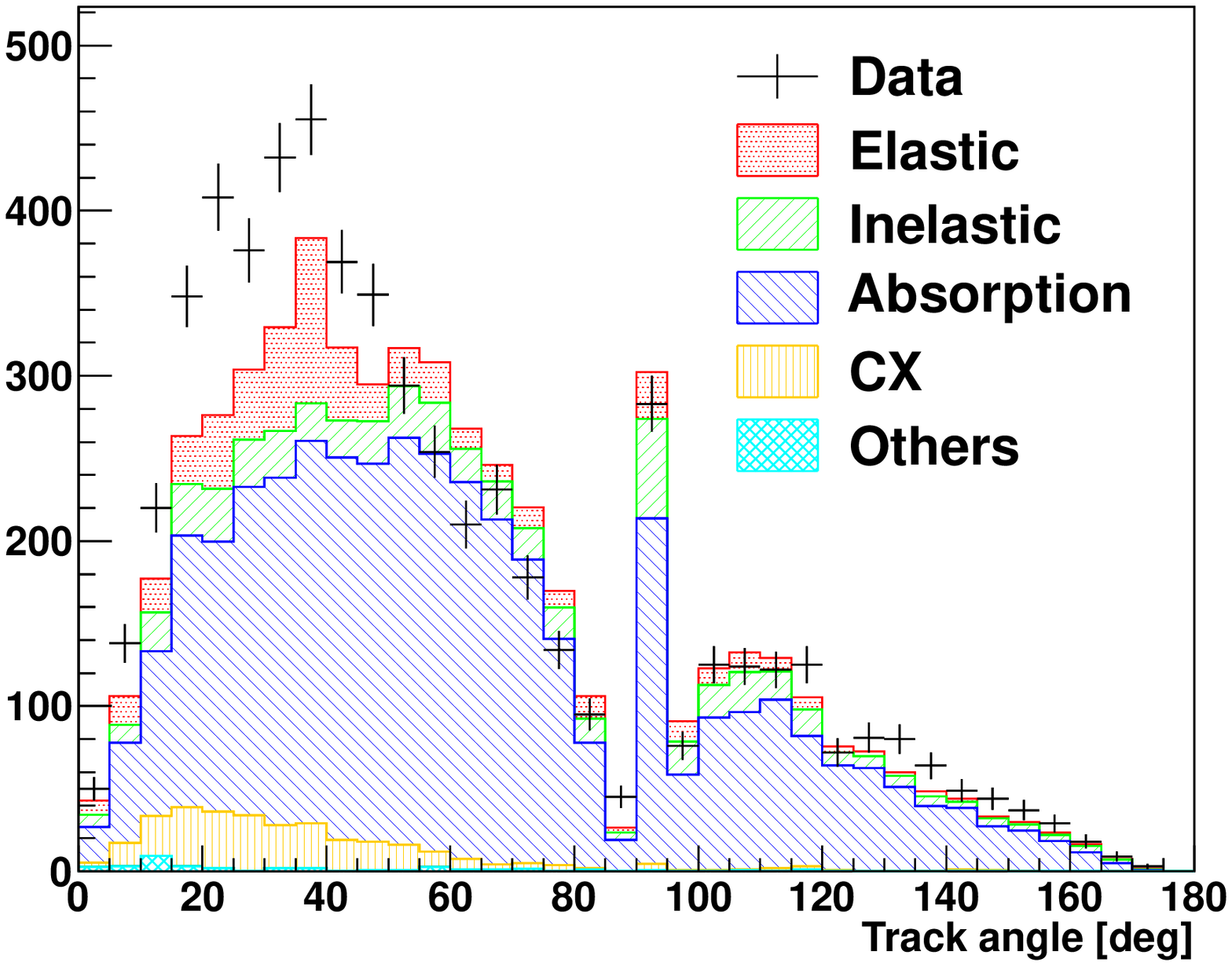}%
    \end{minipage}
    \begin{minipage}{5.9cm}
     \includegraphics[width=6.0cm,clip]{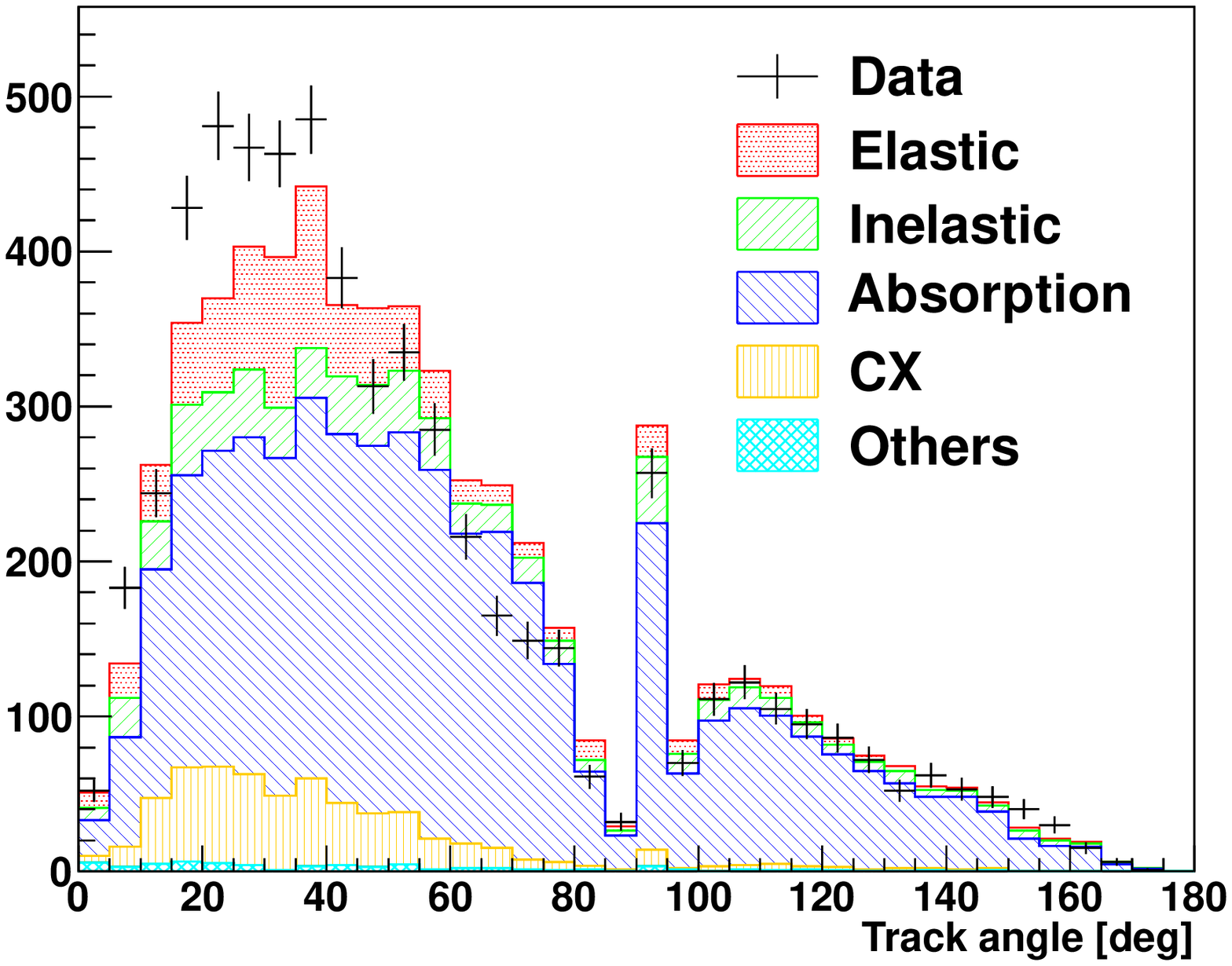}%
    \end{minipage}
    \caption{
    Angular distribution of the reconstructed tracks in the final state 
    for  $p_\pi$ = 201.6 (left), 237.2 (center) and 295.1 (right) MeV/c data, before
    (top) and after (bottom) applying {\it No final $\pi^+$} cut. When the true track
    angle is close to 90 degrees, the track reconstruction algorithm tends
    to reconstruct the track exactly at 90 degrees, so the number of
    events in the bin corresponding to 90 degrees is larger than the
    neighboring bins.
    }
    \label{fig:basic_angle}
   \end{center}
  \end{figure*}

\begin{figure*}[htbp]
   \begin{center}
    \begin{minipage}{5.9cm}
     \includegraphics[width=6.0cm,clip]{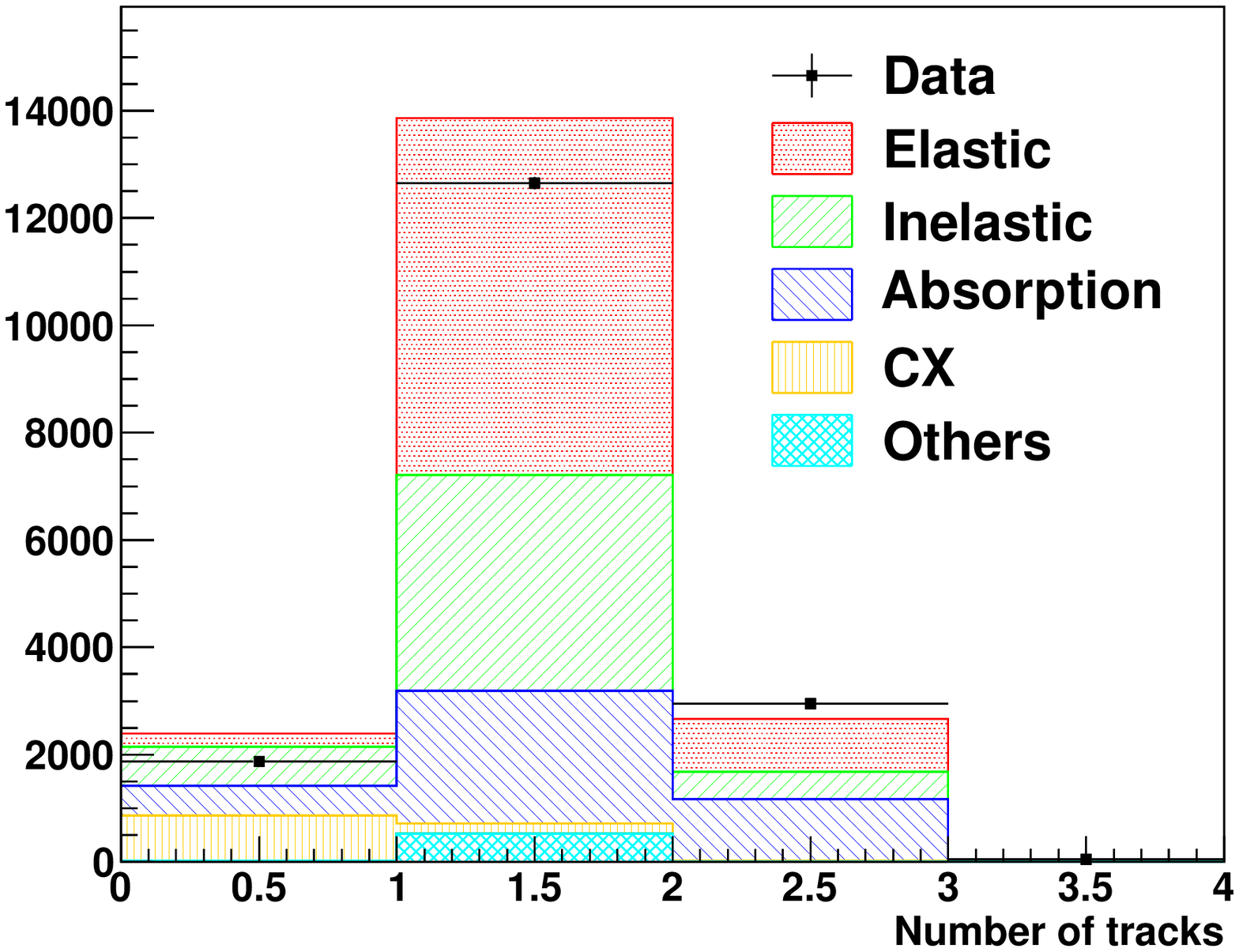}%
    \end{minipage}
    \begin{minipage}{5.9cm}
     \includegraphics[width=6.0cm,clip]{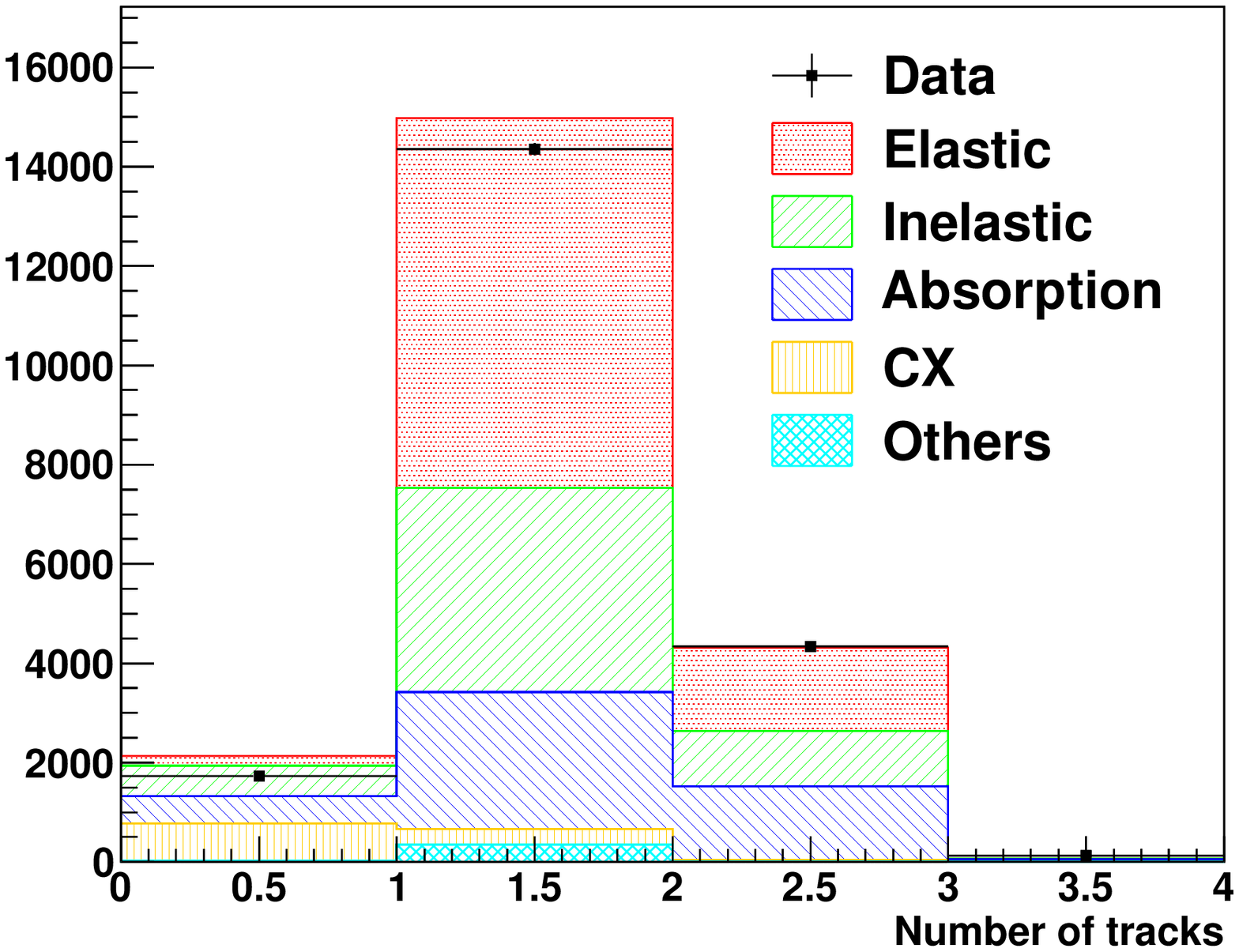}%
    \end{minipage}
    \begin{minipage}{5.9cm}
     \includegraphics[width=6.0cm,clip]{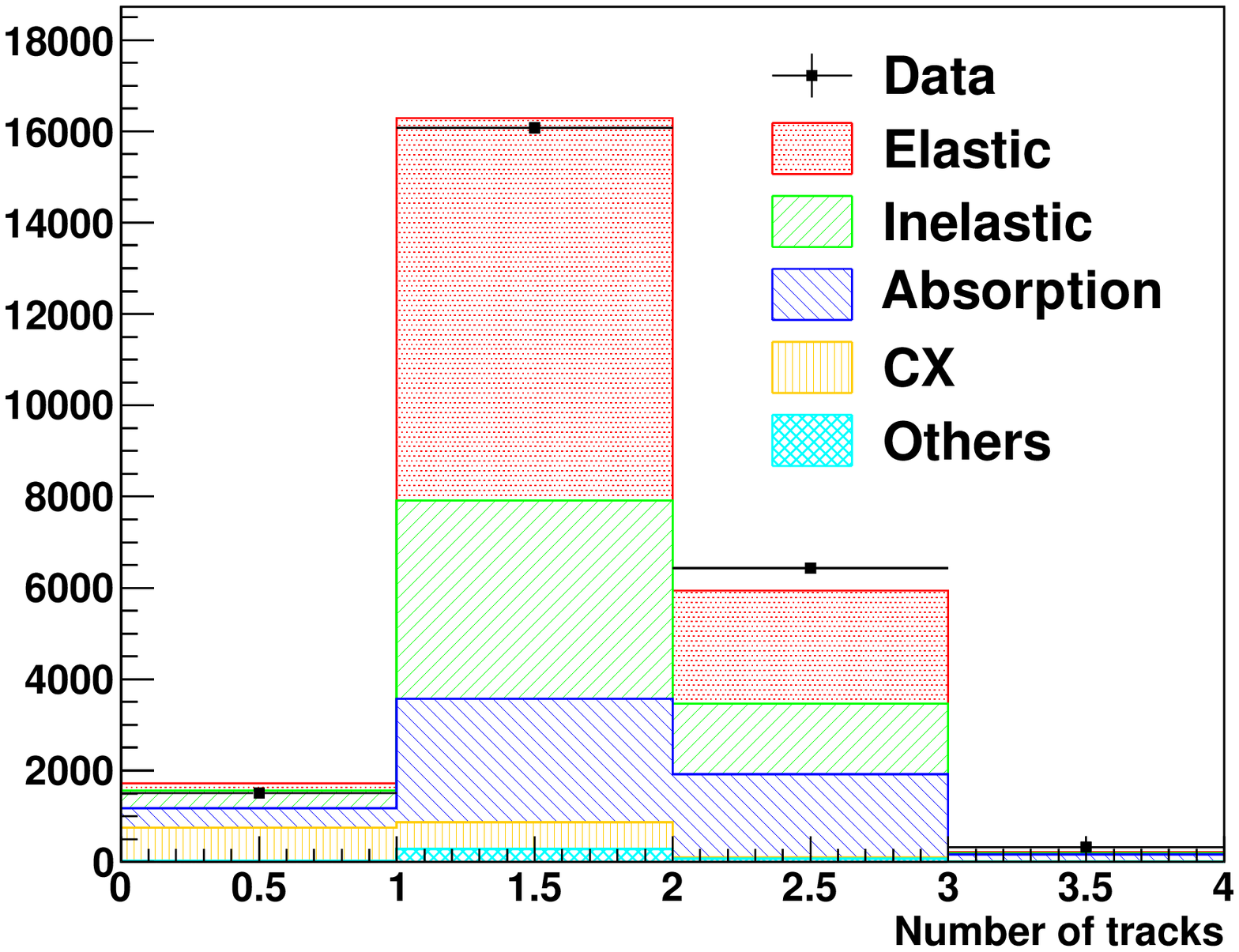}%
    \end{minipage}
    \begin{minipage}{5.9cm}
     \includegraphics[width=6.0cm,clip]{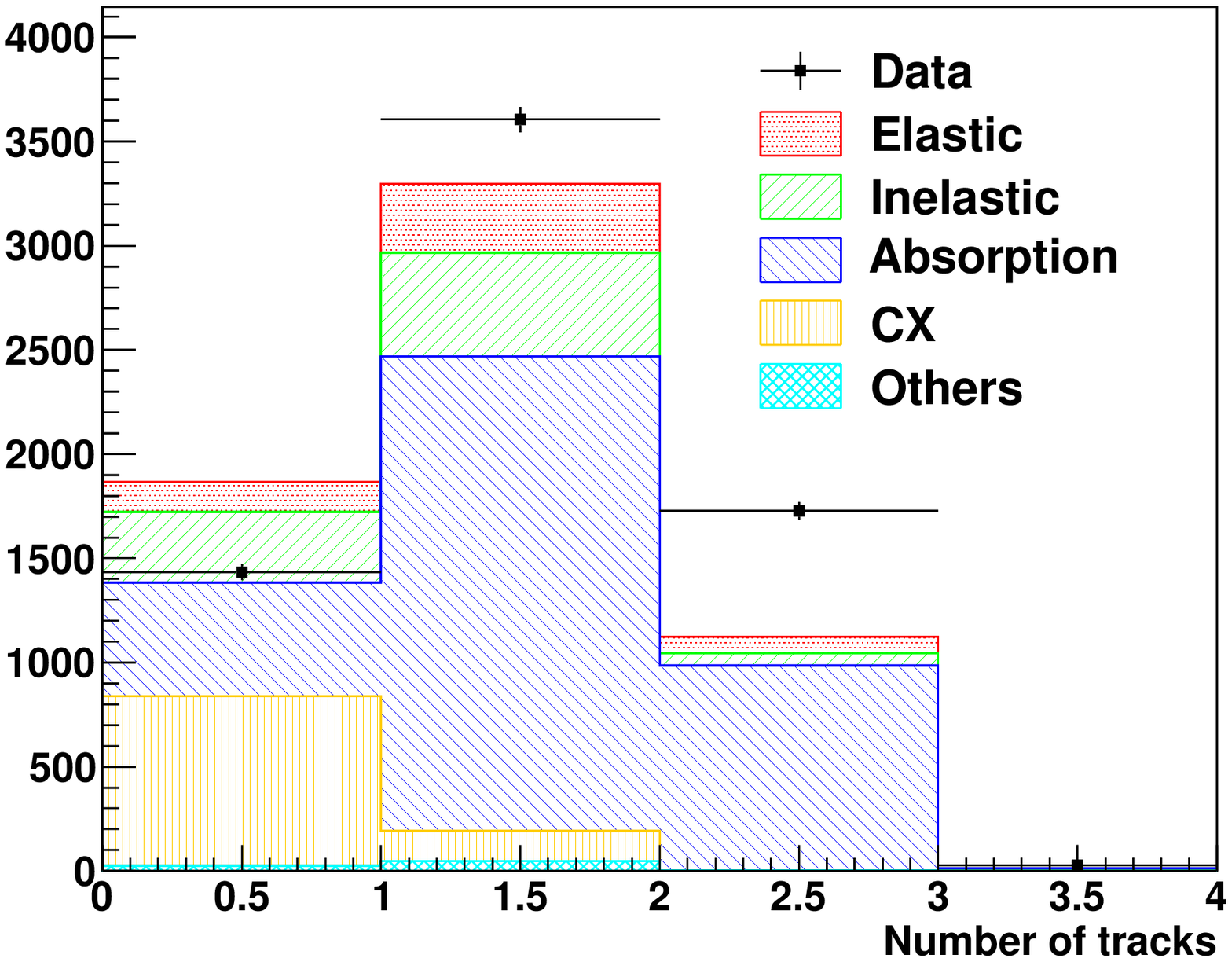}%
    \end{minipage}
    \begin{minipage}{5.9cm}
     \includegraphics[width=6.0cm,clip]{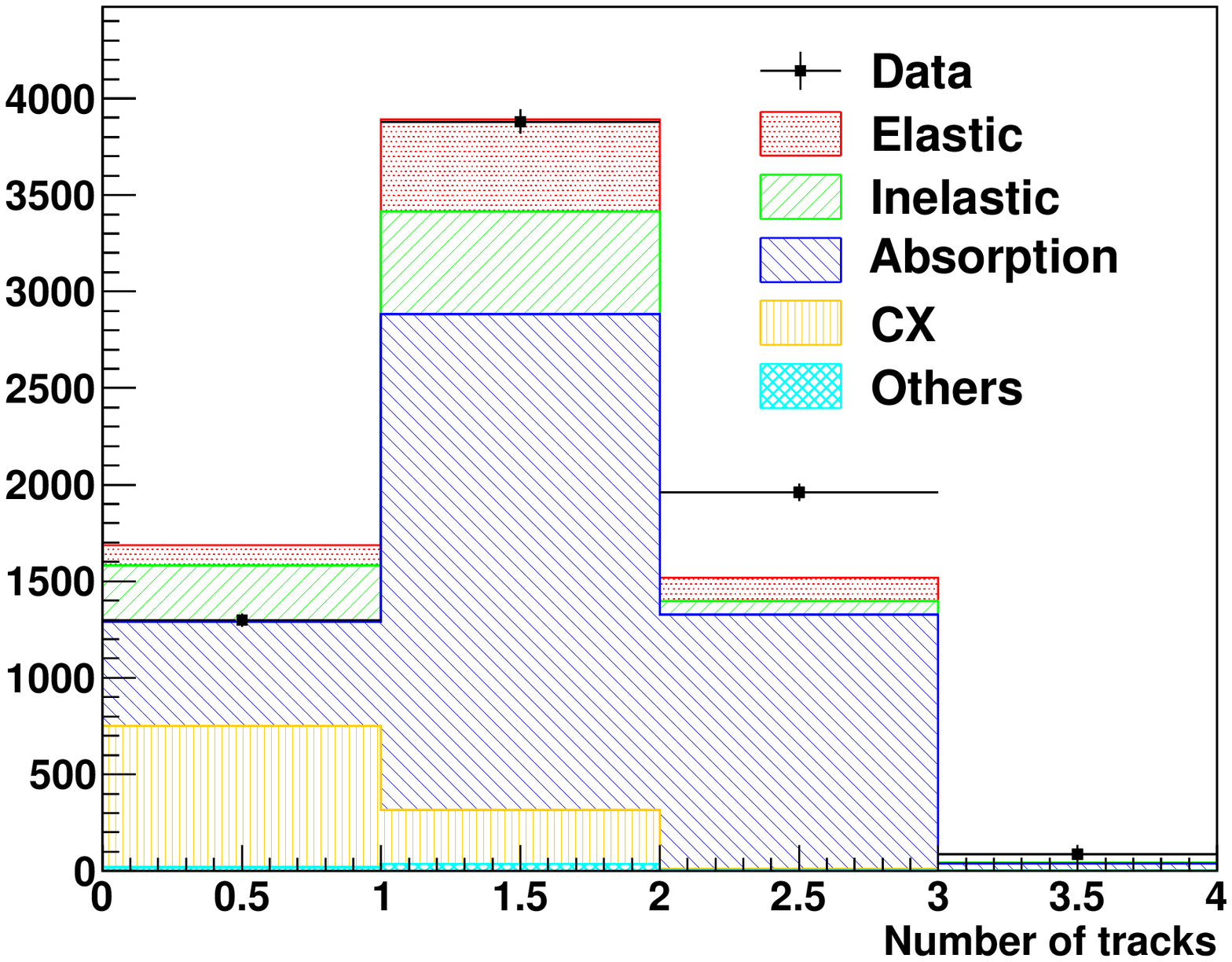}%
    \end{minipage}
    \begin{minipage}{5.9cm}
     \includegraphics[width=6.0cm,clip]{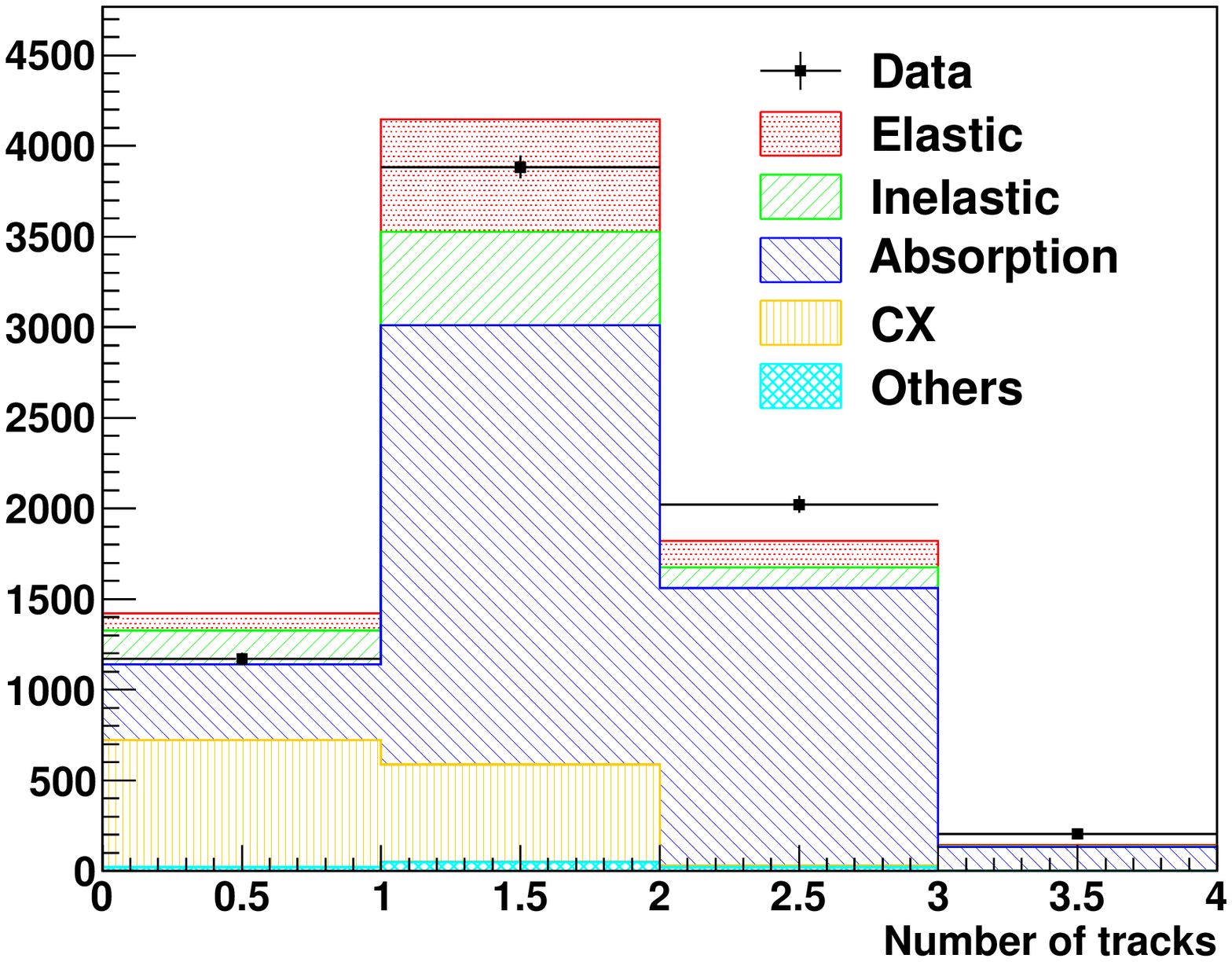}%
    \end{minipage}
    \caption{Distribution of number of
    reconstructed tracks in the final state, for 201.6 (left), 237.2
    (center) and 295.1 (right) MeV/c data sets, before (top) and after
    (bottom) applying {\it No final $\pi^+$} cut.} 
    \label{fig:basic_ntracks}
   \end{center}
  \end{figure*}

  Figure \ref{fig:basic_angle} shows the
  angular distribution of the reconstructed tracks before and after
  applying the {\it No final $\pi^+$} cut, for 201.6, 237.2 and 295.1 MeV/c data sets. In case
  there are multiple tracks in the final state, only the track with the
  smallest value of dQ/dx is selected to fill the histograms in these plots. Figure
  \ref{fig:basic_ntracks} shows the number of
  tracks distribution before and after applying the {\it No final $\pi^+$} cut. After
  applying the {\it No final $\pi^+$} cut, the fraction of ABS and CX events increase,
  and the agreement between data and MC becomes worse. This is
  expected, because the kinematics of the final state particles for ABS
  and CX interactions are not tuned. The event selection efficiency is
  affected by this difference, so it is taken into account in the
  systematic error. 

  
 \clearpage

 \section{Cross-section and error analysis} \label{sec:xsec_calc}
 After the event selection described above, the cross section 
 is obtained by adding the corrections for muon contamination and
 interaction on other nuclei using the following formula,
 
 \begin{equation} \label{eqn:xsec_calc}
   \begin{aligned}
   \sigma_{\mathrm{ABS}+\mathrm{CX}} &= 
    \sigma_{\mathrm{ABS}+\mathrm{CX}}^{pred}
    \times \frac{N_{\mathrm{data}}-N_{\mathrm{BG}}^{pred}}{N_{\mathrm{sig}}^{pred}} \\
    &\times
    \frac{1-R_{\mathrm{TiO}}^{data}}{1-R_{\mathrm{TiO}}^{\mathrm{MC}}}
    \times \frac{1}{1-f_{\mu}}
   \end{aligned}
 \end{equation}

 where $f_{\mu}$ is the fraction of muons in the beam, $R_{\mathrm{TiO}}^{data}$ 
 and $R_{\mathrm{TiO}}^{\mathrm{MC}}$ are the
 fraction of ABS and CX events on Ti or O after the event selection for
 data and MC, shown in Table \ref{tbl:num_xsec_summary}. 
 As mentioned earlier, the outer surface of the fibers has a reflective coating which contains TiO$_2$,
 hence the expected fraction of ABS and CX events in Ti or O in the data must be corrected.
 $R_{\mathrm{TiO}}^{data}$ is estimated from the number of Ti and O
 nuclei (see Table \ref{tbl:fib_nucl}) and the ABS and CX cross-sections for these nuclei, which are
 calculated by interpolating the measured cross sections by a previous external 
 experiment\cite{Ashery}.

  \subsection{Estimation of the systematic errors} \label{sec:syst_detail}
  In this section, we describe in detail the estimation of the systematic errors 
  in the pion interaction measurement, which are summarized in Table~\ref{tbl:syst_summary}. 
  
	\begin{table*}[htbp]
	 \begin{center}
	  \begin{tabular}{lccccc}
	   \noalign{\hrule height 1pt}
	   & \multicolumn{5}{c}{$p_\pi$ at the fiber tracker [MeV/c]} \\
	   & 201.6 & 216.6 & 237.2 & 265.5 & 295.1 \\
	   \hline
	   {\bfseries Systematic errors}              &       &       &       &       & \\
	   ~~Beam profile & 0.9 & 1.2 & 1.0 & 0.6 & 1.2 \\
           ~~Beam momentum & 1.6 & 1.7 & 0.7 & 0.8 & 1.4 \\
           ~~Fiducial Volume & 1.1 & 3.9 & 1.4 & 1.2 & 1.3 \\
           ~~Charge distribution & 2.4 & 2.2 & 2.6 & 2.6 & 2.9 \\
           ~~Crosstalk probability & 0.3 & 0.3 & 0.3 & 0.2 & 0.4 \\
           ~~Layer alignment & 0.5 & 0.8 & 1.1 & 1.0 & 1.4 \\
           ~~Hit efficiency & 0.3 & 0.3 & 0.2 & 0.4 & 0.3 \\
	   ~~Muon contamination	               & 0.5   & 0.8   & 0.9   & 0.3   & 0.2 \\
	   ~~Target material	                       & 0.8   & 0.9   & 0.9   & 0.8   & 1.0 \\
	   ~~Physics models (selection efficiency)     & 2.8   & 4.9   & 2.9   & 4.8   & 3.7 \\
	   ~~\hspace{6.5em} (background prediction) + & 2.8   & 1.8   & 2.4   & 2.3   & 3.3 \\
	   ~~\hspace{17.8em} $-$	               & 6.1   & 3.7   & 3.6   & 1.5   & 1.9 \\
	   \hline
	   ~~Subtotal +	                      & 5.2 & 7.3 & 5.2 & 6.3 & 6.4 \\
           \hspace{4.6em} $-$	                       & 7.5 & 8.0 & 5.9 & 6.0 & 5.8 \\
	   \hline
	   {\bfseries Statistical error (data)}       & 1.7   & 3.1   & 1.7   & 1.8   & 1.7 \\
	   {\bfseries Statistical error (MC)}         & 0.1   & 0.1   & 0.1   & 0.1   & 0.1 \\
	   \hline
	       {\bfseries Total +}	               & 5.5 & 8.0 & 5.5 & 6.6 & 6.7 \\
	       \hspace{2.8em} {\bfseries $-$}	       & 7.7 & 8.6 & 6.2 & 6.3 & 6.1 \\      
	   \noalign{\hrule height 1pt}
	  \end{tabular}
	  \caption{
	  Summary of the statistical and systematic errors in percentage.
	  }
	  \label{tbl:syst_summary}
	 \end{center}
	\end{table*}
  A large part of them are estimated by changing 
  the relevant parameters in the MC. Those systematic errors are defined as
  the difference between the cross section obtained with the nominal MC and
  the changed MC.

   \subsubsection{Beam profile and momentum}
        The properties of the beam are precisely measured in
        through-going pion data by using beam position distribution,
        stopping range distribution and charge distribution. The 
	uncertainty of the momentum is less than 1 MeV/c, and the
        uncertainties on the beam centre position and RMS are $\sim$ 1 mm or less.
        The systematic error for the cross section is evaluated by changing the momentum, the center
        position and the spread of the beam in MC within their uncertainty. 

   \subsubsection{ Fiducial volume }
	An interaction which occurred inside the fiducial volume is
	sometimes reconstructed outside the fiducial volume, or vice versa.
	The fiducial volume systematic error accounts for the
	uncertainty of this effect.
	The size of this effect becomes significant when the definition
	of the FV becomes smaller. Therefore the systematic error is
	estimated by reducing the size of FV by $\sim 20\%$ and
	calculating the difference in the cross section obtained with
	nominal FV and reduced FV.
   \subsubsection{  Charge distribution and crosstalk probability}
        This systematic error is calculated by changing 
	$C_{conv}$, $C_{fluc}$,  $C_{nonlin}$ and the crosstalk probability
	in MC within their uncertainty. 
	The center values and the uncertainties of $C_{conv}$ and
	$C_{nonlin}$ are evaluated by fitting the charge distribution in
	through going pion data obtained at 150 and 295.1 MeV/c settings. The value of
	$C_{fluc}$ is defined from the charge distribution of 1
	p.e. light. The uncertainty of $C_{conv}$, $C_{fluc}$ and
	$C_{nonlin}$ are $\sim 2\%, \sim 6\%$ and $\sim$ 18\%,
	respectively. The crosstalk probability is also estimated by using
	through-going pion data, and the uncertainty is $\sim$ 3\%.
   \subsubsection{   Layer alignment}
	The shift in the position of fiber layers from the nominal
	position is measured using through-going pion data, as
	mentioned in Section \ref{sec:detector_sim}.
	The effect of the uncertainty in the layer position on the cross section
	measurement is estimated by changing the layer position in MC to
	nominal and checking the difference in the measured cross
	section. 

   \subsubsection{   Hit efficiency}
	The efficiency to find a hit above 2.5 p.e.~threshold for the
	charged particles passing through the layer is measured in
	through-going pion data. The efficiency for data was $\sim$
	93\%, while it was $\sim$ 94\% for MC, so the uncertainty is
	assumed to be $\sim$ 1\%. The effect on the cross section is
	estimated by randomly deleting the hits in MC with $\sim$ 1\%
	probability and checking the difference in the resulting cross
	section. 

   \subsubsection{  Muon contamination}
	The uncertainty of muon contamination in the pion beam directly affects the
	normalization of the measured cross section. 
	For the 265.5 and 295.1 MeV/c data sets, the fraction of muons in the beam is
	measured in through-going particle data using CEMBALOS. The absolute error is 
	0.3 and 0.2\%, respectively. For the other data sets, the fraction of muons is
	estimated from the TOF vs. Cherenkov light distribution (Figure \ref{fig:tof_cher}).
	The distribution is projected on the axis perpendicular to the
	threshold line, and the fraction of events above threshold is calculated assuming
	that the distribution follows Gaussian distribution. 
	However, 0.8$\sim$0.9\% of pions which decay just before
	reaching the fiber and identified as incident pion in the event selection.
	Those pions may not be correctly counted with this method.
	Even though the simulation takes into account those pions, 
	to be conservative, 0.8$\sim$0.9\% is assigned for the systematic error.
   \subsubsection{ Target material}
        The number of C, H, O and Ti nuclei in the fiber detector is
        calculated from the dimension and the weight of the fibers.
	The number of C nuclei is estimated to be
	1.518$\pm$0.007$\times$10$^{24}$, and this directly affects 
        the normalization of $\sigma_{\mathrm{ABS}+\mathrm{CX}}$.
        There is also an uncertainty in the number of ABS + CX events on
	O and Ti, which is estimated to be to be 11$\sim$14\% from the
	interpolation of the previous experiment \cite{Ashery}.

   \subsubsection{  Selection efficiency due to physics models}
        The uncertainty in the physics model in MC affects the efficiency to
        select ABS and CX events. This uncertainty corresponds to the uncertainty
        of $N_{\mathrm{sig}}^{pred}$ in Eq. \ref{eqn:xsec_calc}. 
        Table \ref{tbl:syst_summary_effi} summarizes the fractional
	uncertainty of $N_{\mathrm{sig}}^{pred}$ arising from four
sources of uncertainties occurring from the modelling of the physics processes within the MC.

	Each of them are described in the following text.

	\begin{table}[htbp]
	 \begin{center}
	  \begin{tabular}{lccccc}
	   \noalign{\hrule height 1pt}
	   & \multicolumn{5}{c}{$p_\pi$ at the fiber tracker [MeV/c]} \\
	   Error source                               & 201.6 & 216.6 & 237.2 & 265.5 & 295.1 \\
	   \hline
	   Forward / Backward protons                 & 0.4   & 3.2   & 1.4   & 3.2   & 1.8 \\
	   dQ/dx resolution 	                       & 2.7   & 3.6   & 2.2   & 3.4   & 1.7 \\
	   High momentum protons                      & 0.5   & 0.9   & 1.2   & 0.9   & 2.5 \\
	   $\gamma$ conversion	                       & 0.3   & 0.6   & 0.8   & 0.4   & 0.6 \\
	   \hline
	   Subtotal                                   & 2.8   & 4.9   & 2.9   & 4.8   & 3.7 \\
	   \noalign{\hrule height 1pt}
	  \end{tabular}
	  \caption{
	  Summary of the physics model systematic errors related to event
	  selection efficiency (in percentage).
	  }
	  \label{tbl:syst_summary_effi}
	 \end{center}
        \end{table}

	\begin{figure}[htbp]
	 \begin{center}
	  \includegraphics[width=8cm,clip]{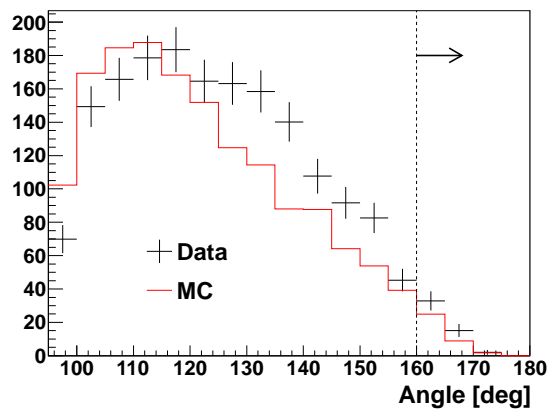}%
	  \caption{%
	  Angular distribution of the backward going proton-like track for data
	  and MC, for ABS and CX events, for $p_\pi =$ 237.2 MeV/c
	  setting. For each event, a proton-track with largest angle is
	  selected and filled in the histogram. The ABS + CX event
	  selection is applied for this plot. The background component
	  (SCAT) is subtracted according to the prediction 
	  in MC, and the histograms are normalized by the number of
	  events after subtraction. 
	  } \label{fig:angle_p_back_datamc}%
	 \end{center}
        \end{figure}

	\begin{figure}[htbp]
	 \begin{center}
	  \includegraphics[width=8cm,clip]{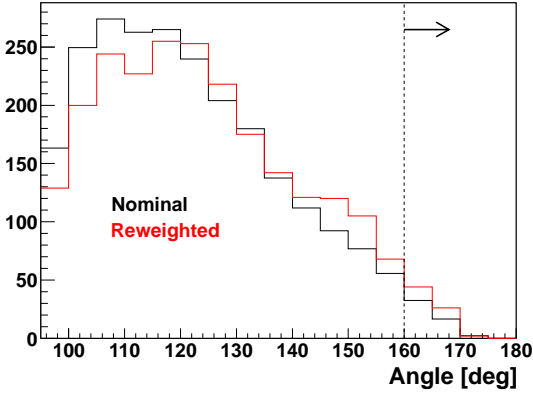}%
	  \caption{%
	  Angular distribution of the backward going proton-like track for
	  nominal and reweighted MC, for ABS and CX
	  events, for $p_\pi =$ 237.2 MeV/c setting. 
	  } \label{fig:angle_p_back_rew}%
	 \end{center}
        \end{figure}

	\smallskip
	{\it  Forward and backward going protons} \\ 

	When a forward going ($\theta<$ 20$^{\circ}$) proton track exists, the position of the
	interaction vertex may be wrongly reconstructed downstream of the actual vertex. 
	When a backward going ($\theta>$ 160$^{\circ}$) proton track exists, the incident track may
	not be identified because it overlaps with the proton track.
	Therefore, the event selection efficiency is affected by the
	fraction of ABS and CX events associated with forward/backward proton tracks.
        
	Figure \ref{fig:angle_p_back_datamc} shows the angular
	distributions for backward going proton-like tracks for data and MC
	with $p_\pi =$ 237.2 MeV/c setting. The data is 1.4 times
	larger in the region above $\theta >$ 160$^{\circ}$. 
	In the case of forward going proton-like tracks the
	data is found to be 1.3 times larger in the region below
	$\theta <$ 20$^{\circ}$. 
	The effect of this difference to the event
	selection efficiency is estimated by using a re-weighted MC
	sample in which the fraction of events with a forward/backward going proton
	track is increased to reproduce the
	data. Figure \ref{fig:angle_p_back_rew} shows the angular distribution of
	proton-like tracks for nominal and re-weighted MC. 
	The event selection efficiency is compared between nominal and
	the re-weighted MC, and the difference is assigned as a systematic 
	error. The error varies from 0.4\% to 3.2\%
	depending on the data sets because the agreement between
	data and MC is different for different data sets.\\

	\begin{figure}[ht]
	 \begin{center}
	  \includegraphics[width=8cm,clip]{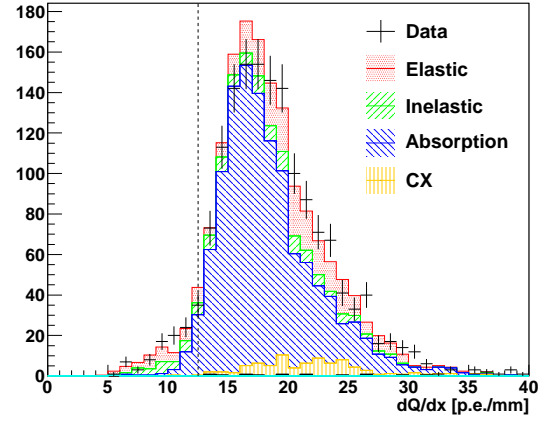}%
	  \caption{
	  Example of dQ/dx distribution for $30^{\circ} < \theta
	  < 60 ^{\circ}$ after event selection, for the projection which was not used for
	  calculating dQ/dx in {\it No final $\pi^+$} cut, for the data set with
	  $p_\pi=$237.2MeV/c setting. The
	  broken line shows the threshold to distinguish pion-like tracks and
	  proton-like tracks. 
	  }\label{fig:dqdx_effi_3060}%
	 \end{center}
        \end{figure}
	

	\smallskip
	{\it dQ/dx resolution} \\

        Events in which a proton track is misidentified as a pion by the dQ/dx cut 
        due to the finite dQ/dx resolution are rejected by the {\it No final  $\pi^+$} cut. 
        The probability to
	misidentify a proton track as a pion track is estimated
	from the probability to pass dQ/dx cut in one projection (U or V)
	but not in the other projection. As mentioned in Section
	\ref{sec:abscx_sel}, the dQ/dx is calculated from U or V
	projection and not from both projections, to minimize the effect
	of saturation of the electronics. Figure \ref{fig:dqdx_effi_3060}
	shows an example of the dQ/dx
	distribution in one projection, when the dQ/dx is required to be
	above threshold in the other projection. 
	The probability to pass the dQ/dx cut in one projection but
	not in the other projection is compared between data and
	MC. For example, in Figure \ref{fig:dqdx_effi_3060}, the fraction
	of events below the threshold is 5\% for data, while it is 4\%
	for MC. Therefore, 25\% error is applied for the number of
	ABS and CX events with proton-like track reconstructed in
	this angular region, at this momentum. 
	Although this error is not small, the
	effect on the total cross section is not significantly large
	because the efficiency of the dQ/dx cut is large ($\sim$
	90\%) and the number of ABS or CX events which do not
	pass this cut is small. \\

	\begin{figure}[ht]
	 \begin{center}
	  \includegraphics[width=80mm]{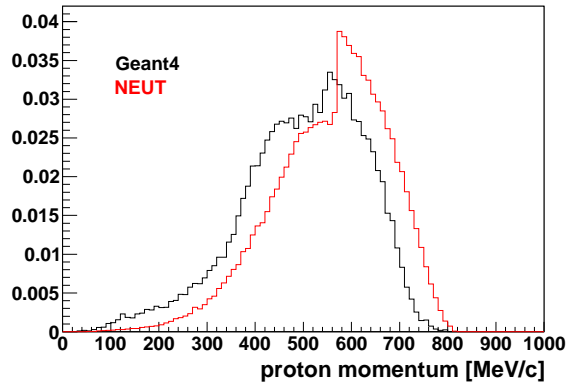}
	  \caption{
	  The predicted momentum distribution of protons from the ABS events, for Geant4
	  (black) and NEUT (red), for $p_\pi = $ 295.1 MeV/c. The histograms
	  are normalized by number of ABS events.
	  }
	  \label{fig:proton_high}
	 \end{center}
        \end{figure}
	
	\smallskip
	{\it High momentum protons} \\

	A small fraction of ABS events have very high momentum protons ($>$
	600 MeV/$c$) in the final state which can not be distinguished
	from pions. Figure \ref{fig:proton_high} shows an example
	of the predicted momentum distribution of protons in the final
	state of ABS events for Geant4 and NEUT, for $p_\pi =$
	295.1 MeV/c case. A large difference
	is observed between two different models and the difference 
	in the fraction of events above 600 MeV/c is assigned as
	the error for the number of high momentum proton events. 
	Because the number of such events is small, the error for those events does not significantly
	affect the error in the cross section.\\

	\smallskip
	{\it Photon conversions} \\

	When the $\gamma$-rays from $\pi^0$ decays in CX events are
	converted to electrons and positrons, these electron
	tracks may be misidentified as pion tracks. These CX events
	are rejected by the {\it No final $\pi^+$} cut. The uncertainty for the
	number of these events are estimated from uncertainty in
	the fraction of CX events and the uncertainty in $\gamma$
	conversion probability. The uncertainty in the fraction of
	CX events is $\sim$ 50\% \cite{Ashery}, and the uncertainty
	of $\gamma$ conversion probability is $\sim$ 5\% \cite{gamma}.
	The systematic errors for the cross section is small because
	the fraction of these events is only $\sim$ 2\% of the
	total number of ABS and CX events.
	
   \subsubsection{ Background estimation from physics models} \label{sec:bg_estimate}
	Pion scattering events are misidentified as ABS and CX, when
	the scattered pion tracks are not identified properly. For
	example, when the pion scattering angle is close to 90 degrees, 
        the pion track may not be reconstructed in one of the two projections, 
        since it may not pass through enough fiber layers.
	Also, due to finite dQ/dx resolution, pion tracks 
	are sometimes misidentified as protons. The tuning based in a linear
	interpolation of data points from the previous measurements does
	not perfectly reproduce the actual cross section. The
	uncertainty for the number of predicted background events is
	estimated in four different categories, as described in the
	following text.\\
	
	\smallskip
	{ \it Pion hadronic scattering} \\

	The number of pion scattering events is compared between data and
	MC in a background enhanced sample. For this data sample, $\pi$-like tracks are
	required in the event selection instead of applying
	{\it No final $\pi^+$} cut. 
	\begin{figure}[ht]
	 \begin{center}
	  \includegraphics[width=80mm]{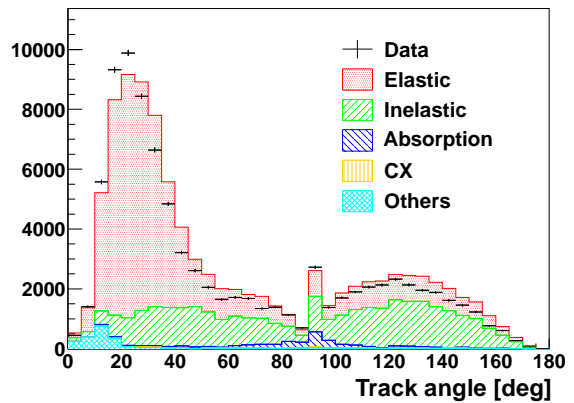}
	  \caption{
	  Angular distribution of $\pi$-like tracks for data and MC with $p_\pi
	  =$ 237.2 MeV/c setting. The histograms are normalized by number of
	  incident pions in data.
	  }
	  \label{fig:angle_bg}
	 \end{center}
        \end{figure}
	Figure \ref{fig:angle_bg} shows the angular
	distribution for $\pi$-like tracks, compared between data and
	MC. The angular	distribution is divided into six different regions: 0-30,
	30-60, 60-100, 80-100, 100-130 and 130-160
	degree. 
	The definition of these are different from
	the angular regions used in the dQ/dx cut
	because the region around 90 degree is important and
	should not be divided into two regions. 
	For each region,
	the difference between data and MC is assigned as the error for
	the number of predicted background events in that region.\\

	\smallskip
	{ \it Back-scattered pions} \\

	For the angular region above 160 degrees, a special data sample
	is prepared in order to compare the difference between data and
	MC. When the scattered angle is near 180 degree, the scattered
	pion track overlaps with the incident pion track. In most cases the
	overlap happens in only one projection, but not in both projections.
	For those back-scattering events, the dQ/dx for the overlapped incident track is large, and
	the scattered pion track is not reconstructed properly in one of
	the two projections.
	\begin{figure}[ht]
	 \begin{center}
	  \includegraphics[width=80mm]{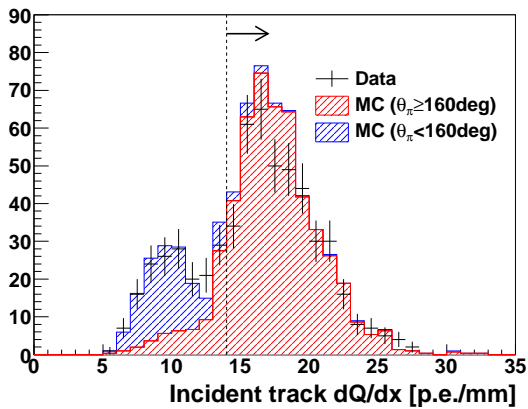}
	  \caption{
	  Example of dQ/dx distribution for incident track for the data set
	  with $p_\pi =$ 237.2 MeV/c setting, after requiring
	  $\pi$-like track in only one of the two projections. The
	  histograms are normalized by number of incident pions. 
	  }
	  \label{fig:dqdx_inc}
	 \end{center}
        \end{figure}
	Figure \ref{fig:dqdx_inc} shows an example of the dQ/dx
	distribution for incident tracks, when a $\pi$-like track (dQ/dx $<$ 15 p.e. / mm) is
	reconstructed in only one projection. 
	The back-scattered pion	data sample is selected by requiring 
	dQ/dx $> 14$ p.e./mm in this plot.
	The difference between data and	MC is assigned as the error for
	the predicted number of back-scattered pion background events. \\

	{ \it Multiple interaction} \\

	Scattered pion tracks may not be
	reconstructed properly when they interact again in
	the fiber tracker. For example, if a pion is
	absorbed right after being scattered, the scattered
	pion track will be too short to be
	reconstructed. Among all of the pion scattering
	events that are misidentified as ABS or CX, $\sim$
	30\% of those are due to multiple interactions like
	this. The uncertainty of the number of events for
	this type of background events is estimated from the
	uncertainty in the cross section from previous
	experiments that we used in MC tuning\cite{Ashery}. 
	For example, for events in which pions are absorbed right after 
        elastic scattering, the uncertainty of the elastic scattering 
        cross section (10\%) and absorption cross section ($\sim$20\%) are applied.
	
	{\it Low momentum pions} \\

\begin{figure}[t]
	 \begin{center}
	  \includegraphics[width=80mm]{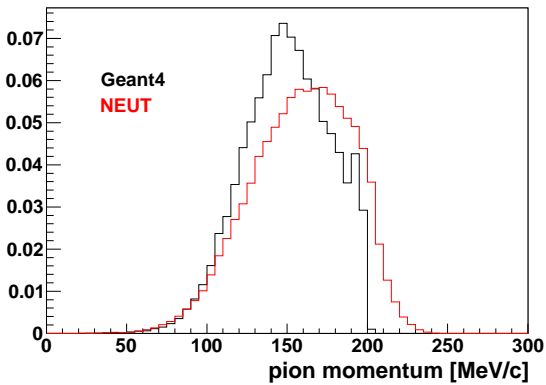}
	  \caption{
	  Predicted momentum distribution of pions from inelastic scattering event, for Geant4
	  (black) and NEUT (red), for $p_\pi = $ 201.6 MeV/c. The histograms
	  are normalized by area.
	  }
	  \label{fig:pion_low}
	 \end{center}
        \end{figure}

	When the momentum of the pions after
	scattering is small ($<$ 130 MeV/c), these pions are
	always identified as protons because the dQ/dx is
	large. 
	Figure \ref{fig:pion_low} shows an example of the
	predicted pion momentum distribution after inelastic
	(quasi-elastic) scattering for Geant4 and NEUT, for
	$p_\pi =$ 201.6 MeV/c. The uncertainty for the number
	of low momentum pion background events is assigned from
	the difference between these two models below 130 MeV/c.


   \subsubsection{ Summary of the systematic errors }
	As summarized in Table \ref{tbl:syst_summary}, the total error is
	$\sim$6.5\%, except for $p_\pi = 216.6$ 
	MeV/c data set, which is roughly half of the errors of the previous
	experiments\cite{Ashery,Navon,navon2}. For $p_\pi = $ 216.6 
	MeV/c data set, the statistical error is relatively large, and the
	systematic error is also found to be large.

  \subsection{Result}
  Table \ref{tbl:num_xsec_summary} summarizes the measurements
  for five momentum data sets. The errors in
  $\sigma_{\mathrm{ABS}+\mathrm{CX}}$ includes both statistical and
  systematic uncertainties. 
  
  \begin{table*}[htbp]
   \hspace{-0.8cm}
   \begin{tabular}{ccccccccc}
    \noalign{\hrule height 1pt}
    $p_\pi$ & \multirow{2}{*}{$N_{\mathrm{data}}$} &
    \multirow{2}{*}{$N_{\mathrm{BG}}^{pred}$} & \multirow{2}{*}{$N_{\mathrm{sig}}^{pred}$} &
    \multirow{2}{*}{$R_{\mathrm{TiO}}^{data}$} & \multirow{2}{*}{$R_{\mathrm{TiO}}^{\mathrm{MC}}$} &
    \multirow{2}{*}{$f_{\mu}$} & $\sigma_{\mathrm{ABS}+\mathrm{CX}}^{pred}$ &
    $\sigma_{\mathrm{ABS}+\mathrm{CX}}$ \\
    ~[MeV/c] & & & & & & & [mbarn] & [mbarn] \\
    \noalign{\hrule height 0.5pt}
    201.6 & 6797 & 1708.9 & 4622.3 & 0.0634 & 0.0808 & 0.0016 & 175.93 & 197.9$^{+10.9}_{-15.3}$ \\
    216.6 & 1814 & 452.3  & 1243.6 & 0.0636 & 0.0731 & 0.0071 & 194.41 & 215.8$^{+17.3}_{-18.6}$ \\
    237.2 & 7671 & 2047.0 & 5572.0 & 0.0624 & 0.0632 & 0.0043 & 214.43 & 216.6$^{+12.0}_{-13.3}$ \\
    265.5 & 6772 & 1851.4 & 5153.7 & 0.0603 & 0.0528 & 0.0054 & 235.92 & 224.8$^{+14.8}_{-14.2}$ \\
    295.1 & 7266 & 1745.4 & 5745.8 & 0.0591 & 0.0518 & 0.0034 & 219.39 & 211.4$^{+14.1}_{-12.9}$ \\
    \noalign{\hrule height 1pt}
   \end{tabular}
   \caption{
   Summary of the measurements. In this table, $p_\pi$ is the momentum of pions at the
   fiber tracker. 
   }
   \label{tbl:num_xsec_summary}
  \end{table*}

\begin{figure}[hb]
   \begin{center}
    \includegraphics[width=80mm]{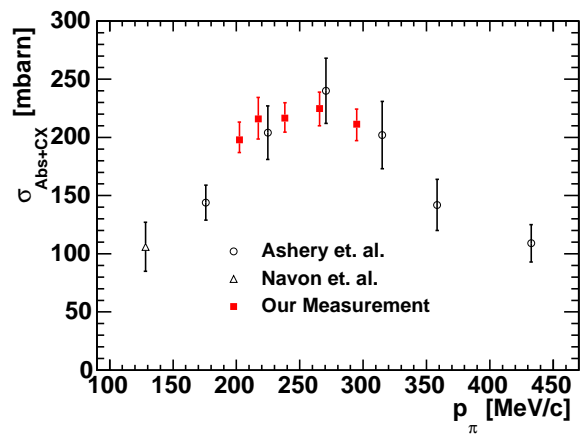}
   \end{center}
   \caption{
   Result of $\sigma_{\mathrm{ABS}+\mathrm{CX}}$ vs. Pion momentum, compared with the
   results from previous experiments.
   }
   \label{fig:xsec_result}
  \end{figure}

  Figure \ref{fig:xsec_result} shows the measured $\sigma_{\mathrm{ABS}+\mathrm{CX}}$
  as a function of pion momentum, compared with the results from previous
  experiments\cite{Ashery,Navon}. 
  
  As already mentioned, the uncertainty in
  our measurement is roughly half of the uncertainty in the previous
  experiments. In these experiments 
  $\sigma_{\mathrm{ABS}+\mathrm{CX}}$ was measured by subtracting the pion
  scattering cross-section from the total cross-section. Since the ABS and CX events were not selected directly there were 
  large errors (typically 5 $\sim$ 10\% in \cite{Ashery}) assigned for
  the subtraction procedure.
  In our measurements, thanks to a fine-grained fully active fiber tracker,
  we were able to measure the ABS + CX interaction directly. \\

  To summarize,  we obtained the cross section for ABS + CX of positive pions 
  in carbon nuclei at an incident momentum between
  201.6 MeV/c to 295.1 MeV/c. The uncertainty  of our measurement is smaller than 
  previous experiments by nearly half due to
  the newly developed fully active scintillation fiber tracker. 
  This result will be a important input to existing models such as Geant4 or NEUT
  to constrain low momentum pion interactions.

\begin{acknowledgments}

We thank for all the technical and financial support received from TRIUMF. 
This work was supported by JSPS KAKENHI Grants Number 22684008, 26247034, 18071005, 20674004
and the Global COE program in Japan. M. I. and K. I. would like to acknowledge support from JSPS.
We acknowledge the support from the NSERC Discovery Grants program, the Canadian Foundation for Innovation’s
Leadership Opportunity Fund, the British Columbia Knowledge Development Fund, and NRC in Canada.
Computations were performed on the GPC supercomputer at the SciNet HPC Consortium.
SciNet is funded by: the Canada Foundation for Innovation under the auspices of Compute Canada; 
the Government of Ontario; Ontario Research Fund - Research Excellence; and the University of Toronto.

\end{acknowledgments}


\nocite{*}

\bibliography{paperNotes}

\end{document}